\documentclass[a4paper,twocolumn,11pt,accepted=2025-07-07]{quantumarticle}
\pdfoutput=1
\usepackage[utf8]{inputenc}
\usepackage[english]{babel}
\usepackage[T1]{fontenc}
\usepackage{tikz}

\usepackage{graphicx}
\usepackage{color}
\usepackage{caption}
\usepackage{braket}
\usepackage{verbatim}
\usepackage{amsmath}
\usepackage{cases}
\usepackage{amsfonts}
\usepackage{mathtools}
\usepackage{braket}
\usepackage{float}
\usepackage{hyperref}
\usepackage{lipsum}
\usepackage{dsfont}
\usepackage{tabularx,ragged2e}
\usepackage{booktabs}
\usepackage{amssymb}
\usepackage{amsthm}
\usepackage[noabbrev]{cleveref}
\usepackage[toc,page,header]{appendix}

\begin{document}

\title{Transforming graph states via Bell state measurements}
\author{Matthias C. L\"{o}bl}
\affiliation{Center for Hybrid Quantum Networks (Hy-Q), The Niels Bohr Institute, University of Copenhagen, Blegdamsvej 17, DK-2100 Copenhagen {\O}, Denmark}
\author{Love A. Pettersson}
\affiliation{Center for Hybrid Quantum Networks (Hy-Q), The Niels Bohr Institute, University of Copenhagen, Blegdamsvej 17, DK-2100 Copenhagen {\O}, Denmark}
\author{Stefano Paesani}
\affiliation{Center for Hybrid Quantum Networks (Hy-Q), The Niels Bohr Institute, University of Copenhagen, Blegdamsvej 17, DK-2100 Copenhagen {\O}, Denmark}
\affiliation{NNF Quantum Computing Programme, Niels Bohr Institute, University of Copenhagen, Blegdamsvej 17, DK-2100 Copenhagen {\O}, Denmark.}
\author{Anders S. S\o{}rensen}
\affiliation{Center for Hybrid Quantum Networks (Hy-Q), The Niels Bohr Institute, University of Copenhagen, Blegdamsvej 17, DK-2100 Copenhagen {\O}, Denmark}

\begin{abstract}
Graph states are key resources for measurement-based quantum computing, which is particularly promising for photonic systems. Fusions are probabilistic Bell state measurements, measuring pairs of parity operators of two qubits. Fusions can be used to connect/entangle different graph states, making them a powerful resource for measurement-based and related fusion-based quantum computing. There are several different graph structures and types of Bell state measurements, yet the associated graph transformations have only been analyzed for specific cases. Here, we provide a full set of graph transformation rules and give an intuitive visualization based on Venn diagrams of local neighborhoods of graph nodes. We derive these graph transformations for all types of rotated type-II fusion, showing that there are five different fusion success cases. Finally, we give application examples of the derived graph transformation rules and show that they can be used to construct graph codes or simulate fusion networks. 
\end{abstract}

\maketitle

\section{Introduction}
Graph states are a class of quantum states that can be represented by a graph where qubits constitute graph nodes and graph edges represent the entanglement structure~\cite{Hein2006}. These states form the key building block for several quantum computing schemes in which computation is performed by measurement operations~\cite{Raussendorf2001, Raussendorf2003,Bartolucci2021}. Bell state measurements are a particular class of two-qubit measurements that measure two parity operators and can be used to engineer such graph states. For photonic qubits, such measurements can be performed with linear optics operations called \textit{fusions}, which successfully implement the two parity measurements with a certain probability~\cite{Browne2005}. A successful fusion can be used to connect different graph states, thus creating larger entangled states and transforming the graph state in a way that single-qubit operations cannot achieve~\cite{Nest2004, Hein2004, Dahlberg2018, Jong2024}. For this reason, fusions have various applications, including quantum repeaters~\cite{Briegel1998,Azuma2015}, generating cluster states~\cite{Kieling2007,GimenoSegovia2015,Lobl2023} for measurement-based quantum computing~\cite{Raussendorf2001}, and fusion-based quantum computing~\cite{Bartolucci2021,Bombin2023}.

Despite the relevance of fusions for such quantum computing/communication architectures, graph transformations have only been investigated for specific cases corresponding to a certain type of two-qubit parity measurement and a restricted class of graph states~\cite{Kok2010,Zaidi2015,Hilaire2023}. First, it was generally assumed that the fusion qubits are unconnected in the graph state and that their neighborhoods do not overlap~\cite{Kok2010,Zaidi2015,Hilaire2023,Patil2023}. Second, the set of parity measurements studied remained incomplete, even when different types of fusion were considered~\cite{Patil2023}. The first assumption often does not hold in fusion networks~\cite{GimenoSegovia2015,Bartolucci2021,Lobl2023} where several fusions are performed and connected fusion qubits or fusion qubits sharing neighbors can result from applying previous fusions. Furthermore, a set of graph transformations that does not cover all fusion types is insufficient for analyzing fusion networks. This is because byproduct gates from previous fusions can rotate the parity measurement corresponding to the present fusion. Thus, a full set of graph transformation rules is required to analyze fusion-based cluster state generation~\cite{GimenoSegovia2015,Lobl2023} or fusion-based quantum computing~\cite{Bartolucci2021} in a graph representation.

Here, we investigate a complete set of graph transformations for all relevant fusions and we also consider the cases where the fusion qubits are connected\footnote{Graph transformations for single-qubit measurements can be found in Refs.~\cite{Hein2004, Dahlberg2018} and can be expressed by local graph complementations~\cite{Nest2004, Cabello2011, Adcock2020} and vertex deletion (vertex minors).}. Upon success, the considered fusions correspond to simultaneous measurements of two weight-two operators (parities) from the Pauli group. Measurements of this type are of particular interest as they keep the considered state a stabilizer state~\cite{Gottesman1998}. Therefore, our graph transformations can also be used to treat the so-called stabilizer states~\cite{Anders2006} because they are locally equivalent to graph states~\cite{Nest2004, Schlingemann2001}. We find that there are five different cases of such parity measurements, and we derive graph transformations for all these cases. Knowing the graph transformations enables simulating fusion networks in a sparse graph representation instead of a representation based on stabilizer tableau matrices~\cite{Gottesman1998, Hein2006}. Furthermore, we provide an intuitive visualization of the graph transformations using Venn diagrams. This visualization makes it easy to apply the graph transformations by hand and find protocols to generate resource graph states with fusions. We give an example of a construction of an eight-qubit graph code from a linear chain graph state that we found manually using the graph transformations~\cite{Bell2022}. Finally, the graph transformations are also highly useful for automatically searching optimized graph-state generation protocols~\cite{Lobl2024}, and we provide an implementation in Python~\cite{git2023}.

The manuscript is organized as follows: In Section~\ref{sec:parity}, we introduce the different parity measurements that correspond to a successful fusion. In Section~\ref{sec:trafo}, we introduce graph states and the graph transformations that are caused by the parity measurements. In Section~\ref{sec:appl} we discuss fusion networks and the construction of graph codes as potential applications of the derived graph transformation rules.

\begin{figure}[!t]
\includegraphics[width=1.0\columnwidth]{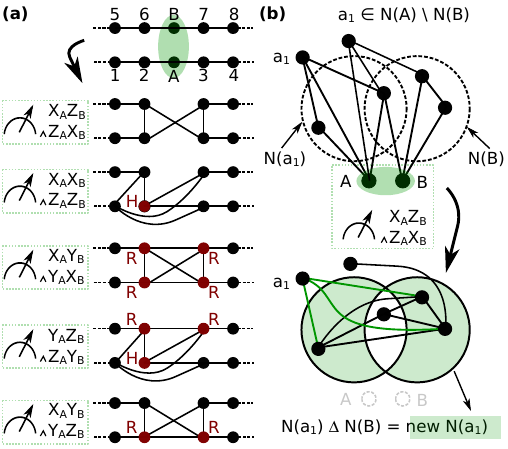}
\caption{\label{fig:intro}\textbf{(a)} Fusing linear chain graph states by different parity measurements corresponding to a successful fusion. The state after the fusion is local Clifford equivalent to a graph state, with the local Clifford gates $H=\frac{1}{\sqrt{2}}\big(\begin{smallmatrix}1 & 1\\1 & -1\end{smallmatrix}\big)$, $R=\big(\begin{smallmatrix}1 & 0\\0 & i\end{smallmatrix}\big)$ shown in red. Note that there is some freedom of choice for these local gates which can lead to a breaking of the symmetry when representing the state as a graph. This is, for instance, the case for the fusion corresponding to the measurement of the two parities $X_AX_B\land Z_AZ_B$ where the symmetry among the qubits is broken by the choice of qubit 2 as the so-called special neighbor. Furthermore, the measurement outcomes may give rise to additional Pauli $Z$ gates that correspond to stabilizer signs. \textbf{(b)} Graph transformation upon measuring the parities $X_AZ_B\land Z_AX_B$. In the original graph state, qubit $a_1$ is connected to the fusion qubit $A$, not to the fusion qubit $B$ (the case $a_i\in N(A)\setminus N(B)$ in Eq.~\eqref{eq:xzzx_example_unconn_0} where $N(A)$ represents the neighborhood of qubit $A$). The new neighborhood of qubit $a_1$ is given by the so-called symmetric difference of its old neighborhood $N(a_1)$ and the neighborhood $N(B)$ of the fusion qubit $B$: $N(a_1)\Delta N(B)=\left(N(a_1)\setminus N(B)\right)\cup\left(N(B)\setminus N(a_1)\right)$. The symmetric difference representing the new qubit neighborhood is illustrated by the green area in the Venn diagram.}
\end{figure}

\section{The different parity measurements}
\label{sec:parity}
We first investigate the number of different Bell-type measurements corresponding to products of Pauli operators\footnote{Representing the $Z$-eigenstates $\ket{0}$ and $\ket{1}$ as vectors $\big(\begin{smallmatrix}1 \\0\end{smallmatrix}\big)$ and $\big(\begin{smallmatrix}0 \\1\end{smallmatrix}\big)$, respectively, we define the Pauli matrices as $X=\big(\begin{smallmatrix}0 & 1\\1 & 0\end{smallmatrix}\big)$, $Y=\big(\begin{smallmatrix}0 & -i\\i & 0\end{smallmatrix}\big)$, $Z=\big(\begin{smallmatrix}1 & 0\\0 & -1\end{smallmatrix}\big)$}. These measurements correspond to simultaneously measuring two pairs of parities between two qubits $A, B$ resulting in two bits of information. For photonic qubits, such a measurement is realized by a successful type-II fusion which destructively measures the two qubits~\cite{Browne2005} (see Appendices~\ref{sec:setup},~\ref{sec:circuit}). With standard fusion circuits, the success of the type-II fusion is probabilistic, meaning that the two parities are typically measured with a success probability of $p_s=0.5$, yet we show in Appendix~\ref{sec:probability} that there are also boundary cases in which the fusion is deterministic. The success probability can be increased by boosting with additional photons~\cite{Grice2011,Ewert2014}. A successful fusion results in new connections between previously detached components of a graph state~\cite{Browne2005,Kok2010}. In contrast, fusion failure corresponds to two independent single-qubit measurements~\cite{Bartolucci2021} (see Appendix Fig.~\ref{fig:circiut}(a)) and does not make such a connection. The standard fusion measures the two stabilizers of the Bell states, the parities $X_AX_B\land Z_AZ_B$, upon success (see Appendix~\ref{sec:setup} for a setup). The measured parities can be rotated by applying single-qubit gates before the standard fusion~\cite{Zaidi2015,GimenoSegovia2015,Lobl2023}. Here, we consider gates from the so-called local Clifford group~\cite{Calderbank1998,Gottesman1998} that is generated by the gates $H=\frac{1}{\sqrt{2}}(\ket{0}\bra{0}+\ket{0}\bra{1}+\ket{1}\bra{0}-\ket{1}\bra{1}), R=\ket{0}\bra{0}-i\ket{1}\bra{1}$ and which normalizes the Pauli group. Therefore, all the measured operators remain part of the Pauli group, resp. the considered state remains a stabilizer state. When a fusion between two qubits $A$ and $B$ is successful, the measured parities can thus be expressed as:
\begin{align}
    &\sigma_A^{(1)}\otimes\sigma_B^{(2)}\notag\\
    \land\, &\sigma_A^{(3)}\otimes\sigma_B^{(4)}
\end{align}
where $\sigma_i^{(j)}\in\{X,Y,Z\}$ is from the Pauli group and $\otimes$ denotes the tensor product, which we will not write explicitly in the following. Note that $\sigma_A^{(1)}\neq\sigma_A^{(3)}$ and $\sigma_B^{(2)}\neq\sigma_B^{(4)}$ must be fulfilled. Otherwise, the two parities would either be identical, or they would not commute and therefore could not be measured simultaneously. This gives $3\cdot3\cdot2\cdot2=36$ possible combinations of stabilizers. The order of the stabilizers does not matter, resp. swapping the parity $\sigma_A^{(1)}\sigma_B^{(2)}$ with the parity $\sigma_A^{(3)}\sigma_B^{(4)}$ is the same measurement. Furthermore, parity measurements such as $X_AX_B\land Z_AZ_B$ and $X_AX_B\land Y_AY_B$ are identical because $Y_AY_B = (X_AX_B) \cdot (Z_AZ_B)$. As a consequence, the $36$ pairs of parities can be partitioned into six sets of the same size\footnote{All sets of parity measurements can be found in our Python implementation of the graph transformations~\cite{git2023}}. Each set contains equivalent pairs of parity measurements, and the pairs of parity measurements from two different sets are not equivalent. We select one pair of parities per set to represent the measurement. These pairs of parities are shown in Table~\ref{tab:fusions}.
\begin{table}
\begin{center}
 \begin{tabular}{ |c||c|c|c|c|c|c| } 
 \hline
 $\sigma^{(1)}_A\sigma^{(2)}_B$ & $XX$ & $XY$ & $XZ$ & $YZ$ & $XY$ & $YX$\\ 
 $\sigma^{(3)}_A\sigma^{(4)}_B$ & $ZZ$ & $YX$ & $ZX$ & $ZY$ & $YZ$ & $ZY$\\ 
 \hline
\end{tabular}
\caption{\label{tab:fusions}Parity measurements upon fusion success.}
\end{center}
\end{table}

Furthermore, the labeling of the fusion qubits, $\sigma_A^{(1)}\sigma_B^{(2)}\land\sigma_A^{(3)}\sigma_B^{(4)}$ or $\sigma_A^{(2)}\sigma_B^{(1)}\land\sigma_A^{(4)}\sigma_B^{(3)}$, does not matter for the derivation of the graph transformation rules. Therefore, we consider the two last cases in Table~\ref{tab:fusions} as one case, as the parity pairs are identical when swapping the labels $A, B$. The remaining five cases cover all simultaneous two-parity measurements from the Pauli group. Setups and circuits that implement these parity measurements (upon fusion success) are given in the Appendices~\ref{sec:setup},~\ref{sec:circuit}. Note that there are different setups to implement the same parity measurements. Even if two setups give the same measurement upon fusion success, the measurement upon fusion failure can differ (see Appendix~\ref{sec:setup}). All fusion failures correspond to single-qubit Pauli measurements, and the corresponding graph transformations can be found in Ref.~\cite{Hein2006}.

\section{Graph transformations}
\label{sec:trafo}
In this section, we will derive the graph state transformation rules for some of the five different cases in Table~\ref{tab:fusions}. With the help of such rules, one can find constructions of specific graph states via fusions. In Fig.~\ref{fig:intro}(a), we illustrate how the different parity measurements lead to different graph transformations. To this end, we chose an example consisting of two linear chain graph states. The different two-qubit parity measurements lead to different graph states that are not local Clifford equivalent~\cite{Cabello2011,Adcock2020} even though the fusion types all fall into the same class of entangled measurements (Bell state measurements)~\cite{DelSanto2023}. Fig.~\ref{fig:intro}(b) illustrates how graph transformations can be visualized by Venn diagrams.

We start with a definition of graph states and stabilizer states. Assume a graph $G(V, E)$ where every node in $V$ represents a qubit. A graph state can be constructively defined as $\ket{G}=\prod_{(i,j)\in E} CZ_{ij}\ket{+}^{\otimes{|V|}}$ where $CZ_{ij}=\ket{0_i0_j}\bra{0_i0_j}+\ket{0_i1_j}\bra{0_i1_j}+\ket{1_i0_j}\bra{1_i0_j}-\ket{1_i1_j}\bra{1_i1_j}$ represents an entangling controlled-$Z$ gate between qubits $i, j$ and $\ket{+}=\frac{1}{\sqrt{2}}\left(\ket{0}+\ket{1}\right)$~\cite{Hein2006}. The edges of the graph, $E$, thus represent the entanglement structure and the graph state is well-defined as all the gates $CZ_{ij}$ commute. Graph states are part of the larger class of stabilizer states~\cite{Gottesman1998}. This formalism represents a stabilizer state by a set $\mathcal{S}$ of independent and commuting operators from the Pauli group, the stabilizer generators, fulfilling $\forall S\in\mathcal{S}: S\ket{G}=+1\ket{G}$. Every stabilizer state is locally equivalent to a graph state~\cite{Schlingemann2001, Nest2004} meaning that it can be obtained from a graph state by applying local gates from the Clifford group~\cite{Calderbank1998,Gottesman1998}. The stabilizer generators of a graph state can be represented by the set of all operators of the form $X_i\prod_{j\in N(i)}Z_j$~\cite{Raussendorf2003,Hein2006}. Here, $N(i)$ represents the graph neighborhood of qubit $i$ that constitutes all qubits $j$ for which there is an entangling controlled-$Z$ gate $CZ_{ij}$. Defining a graph state in terms of its stabilizers or its edges is equivalent.

We initially consider the case where the two fusion qubits, labeled $A, B$ in the following, are unconnected and write the stabilizer generators before the fusion. First, there are two stabilizer generators $S_A$, $S_B$ that have a Pauli $X$-operator on the fusion qubits $A, B$. Second, we distinguish four different classes of qubits resp. stabilizer generators: If qubit $j$ is connected to $A$ but not $B$ there is a stabilizer generator containing $X_jZ_A$ (case 1). If $j$ is connected to $B$ but not $A$, the corresponding stabilizer generator contains the term $X_jZ_B$ (case 2). If $j$ is connected to both $A, B$, there is a stabilizer generator containing the term $X_jZ_BZ_A$ (case 3). Finally, some qubits are connected to neither $A, B$ corresponding to stabilizer generators with no support on $A, B$ (case 4). Listing all these cases yields\footnote{To represent the same single-qubit operator $\sigma$ acting on a set of qubits $Q$, we use the convention $\sigma_Q:=\prod_{q\in Q}^{\otimes}\sigma_q$}:
\begin{align}
    &S_A = Z_{N(A)} X_A \label{eq_stabs_all_1}\\
    &S_B = Z_{N(B)} X_B\label{eq_stabs_all_2}\\
    &\forall a_i \in N(A)\setminus N(B): X_{a_i}Z_{N(a_i)\setminus A}Z_A\label{eq_stabs_all_3}\\
    &\forall b_i \in N(B)\setminus N(A): X_{b_i}Z_{N(b_i)\setminus B}Z_B\label{eq_stabs_all_4}\\
    &\forall c_i \in N(A)\cap N(B): X_{c_i}Z_{N(c_i)\setminus A \setminus B}Z_AZ_B\label{eq_stabs_all_5}\\
    &\forall d_i \notin N(B)\cup N(A): X_{d_i}Z_{N(d_i)}\label{eq_stabs_all_6}
\end{align}

The stabilizers of the initial graph state are different when the two fusion qubits $A$, $B$ are connected, and we treat this case independently in Appendix~\ref{sec:fuse_connected}. Furthermore, note that even for unconnected fusion qubits, not all terms in \cref{eq_stabs_all_3,eq_stabs_all_4,eq_stabs_all_5,eq_stabs_all_6} always exist. For example, stabilizers of the form $S_{a_i}$ could be missing when the entire neighborhood of qubit $A$ is a subset of the neighborhood of qubit $B$. To derive rules for the graph transformations, some of these cases have to be treated separately.

We derive all the graph transformation rules in two main steps. First, we determine how the stabilizers of the graph state are modified by the parity measurements~\cite{Gottesman1998,Aaronson2004,Fowler2009}. As the resulting state is not necessarily a graph state, we determine a local Clifford equivalent graph state and the required single-qubit Clifford gates to bring it into the form of a graph state in the second step~\cite{Nest2004}\footnote{We do not consider the stabilizer signs as those also depend on the probabilistic measurement result of the fusion. The provided graph transformations, together with the local Clifford gates, therefore give a graph basis state~\cite{Looi2008} that is equivalent to a graph state up to local Pauli $Z$-gates.}.

Modifying the stabilizer generators upon fusion is done similarly to Ref.~\cite{Fowler2009}. When a measurement is performed, all stabilizer generators that anti-commute with one of the two measured parities are not measurable simultaneously. A maximum set of simultaneously measurable stabilizer generators is obtained in the following way: multiply one stabilizer generator $S_i$ that anticommutes with the first measured parity on all the other anti-commuting stabilizer generators making them commute with the first parity. Ignoring $S_i$, repeat the procedure with a stabilizer $S_j$ that anticommutes with the second parity measured by the fusion. The stabilizer generators $S_i, S_j$ are then removed and, in the procedure from Ref.~\cite{Fowler2009}, the measured operators would be added as new stabilizers. In this case, the support on the fusion qubits $A, B$ could be removed from all other stabilizer generators by multiplication of stabilizer generators (essentially because a measurement detaches the measured qubits from the rest of the state). The Bell state measurement is destructive for photonic qubits and they therefore cannot be measured again. After dropping the $S_j$ and $S_i$, we therefore simply remove the Pauli matrices acting on the fusion qubits from all other stabilizer generators instead of adding the measured parities as new stabilizers.\footnote{For single-qubit measurements~\cite{Hein2004,Hein2006} the procedure is analogous (see Appendix~\ref{sec:single_measure}).}.

A measured parity may commute with all stabilizer generators of the graph state. In this case, the measured parity is either a stabilizer itself or, if not, the state after the measurement is obtained by adding the measured parity to the list of stabilizer generators~\cite{Fowler2009}. As the number of stabilizer generators of a pure state is equal to the number of qubits, the commuting parity must be a stabilizer in our case\footnote{This case can correspond to a deterministic fusion (see Appendix~\ref{sec:probability}).}. All that needs to be done is to remove the Pauli matrices acting on the destructively measured fusion qubits from the stabilizer generators and determine a new set of independent stabilizer generators.

In the following, we illustrate the procedure by considering two different types of fusion corresponding to the pairs of parity measurements $X_AZ_B\land Z_AX_B$ and $X_AX_B\land Z_AZ_B$. All other cases for unconnected fusion qubits are considered in Appendix~\ref{sec:fuse_connected}, while all cases for connected fusion qubits are treated in Appendix~\ref{sec:fuse_unconnected}.

\subsection{$X_AZ_B\land Z_AX_B$ ($A$, $B$ detached)}
\begin{figure*}
\includegraphics[width=1.0\textwidth]{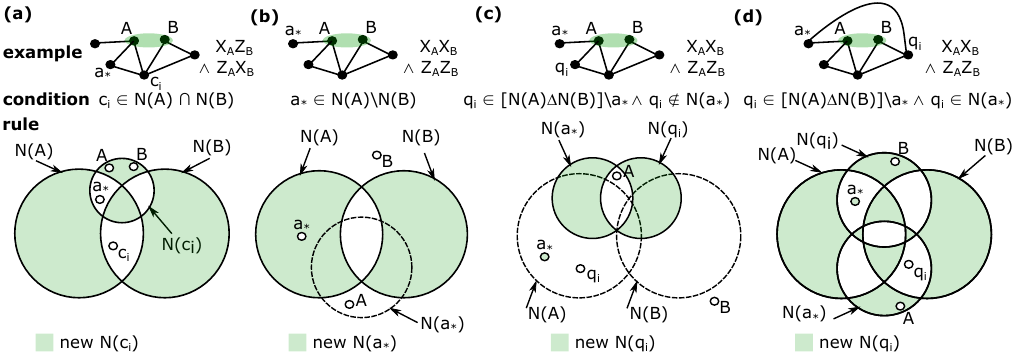}
\caption{\label{fig:Venn_main} Venn diagrams representing the graph transformations for different fusion types when the two fusion qubits $A$ and $B$ are not connected. \textbf{(a)} Graph transformation upon measuring the parities $X_AZ_B\land Z_AX_B$ for a qubit $c_i$ that is connected to both fusion qubits $A, B$. The Venn diagram represents Eq.~\eqref{eq:xzzx_example_unconn_2} (where the $a_*$ symbol in the figure can be ignored). The depicted Venn diagram also represents the $X_AX_B\land Z_AZ_B$ fusion for the case $c_i \in N(a_*)$ in Eq.~\eqref{eq_xxzz_unconn_2}. \textbf{(b)} The new neighborhood of the special neighbor qubit $a_*$ when doing the parity measurement $X_AX_B\land Z_AZ_B$. This Venn diagram corresponds to Eq.~\eqref{eq_xxzz_unconn_a_st}. \textbf{(c)} Venn diagram representing the neighborhood transformation for the case $q_i\notin N(a_*)$ in Eq.~\eqref{eq_xxzz_unconn_0}. \textbf{(d)} Venn diagram for the $X_AX_B\land Z_AZ_B$ fusion in the case $q_i \in N(a_*)$ in Eq.~\eqref{eq_xxzz_unconn_0}. We show an example for a qubit in the neighborhood of fusion qubit $B$ ($q_i\in N(B)\setminus N(A)$) but the same rule would apply for $q_i\in N(A)\setminus N(B)$.}
\end{figure*}

The simultaneous measurement of the parities $X_AZ_B$ and $Z_AX_B$ has been considered in Refs.~\cite{Kok2010,Zaidi2015,Patil2023}, but only for the special case where the neighborhoods of the fusion qubits are not overlapping ($N(A)\cap N(B)=\emptyset$) and the fusion qubits are not connected ($A\notin N(B)$). In the following, we consider the parity measurement $X_AZ_B\land Z_AX_B$ for unconnected fusion qubits without any further assumptions on the graph structure, while connected fusion qubits are treated in Appendix~\ref{sec:fuse_connected}.

The measured parity $X_AZ_B$ anti-commutes with the stabilizers in ~\cref{eq_stabs_all_2,eq_stabs_all_3,eq_stabs_all_5} and we therefore multiply Eq.~\eqref{eq_stabs_all_2} on the stabilizers in ~\cref{eq_stabs_all_3,eq_stabs_all_5}. $Z_AX_B$ anti-commutes with the stabilizers in ~\cref{eq_stabs_all_1,eq_stabs_all_4,eq_stabs_all_5} (also after multiplication with Eq.~\eqref{eq_stabs_all_2} from treating the first parity). Therefore, we multiply all stabilizers in ~\cref{eq_stabs_all_4,eq_stabs_all_5} with Eq.~\eqref{eq_stabs_all_1}. As described before, we drop the Pauli matrices acting on the measured qubits, which yields the following stabilizer generators:
\begin{flalign}
    &\forall a_i \in N(A)\setminus N(B): X_{a_i}Z_{N(a_i)\Delta N(B)}\label{eq:xzzx_example_unconn_0}\\
    &\forall b_i \in N(B)\setminus N(A): X_{b_i}Z_{N(b_i)\Delta N(A)}\label{eq:xzzx_example_unconn_1}\\
    &\forall c_i \in N(A)\cap N(B):\notag\\
    &\hspace{0.7cm}X_{c_i}Z_{N(c_i)\Delta N(A)\Delta N(B)}\label{eq:xzzx_example_unconn_2}\\
    &\forall d_i \notin N(B)\cup N(A): X_{d_i}Z_{N(d_i)}\label{eq:xzzx_example_unconn_3}
\end{flalign}
Here $S_1\Delta S_2$ denotes the symmetric difference between the two sets $S_1$ and $S_2$\footnote{For two sets $S_1, S_2$, the symmetric difference is $S_1\Delta S_2:=\left(S_1\setminus S_2\right)\cup\left(S_2\setminus S_1\right)$, i.e. the symmetric difference consists of the elements which are in one of the sets and not the other. When $S_1, S_2$ represent sets of qubits, $Z_{S_1}\cdot Z_{S_2}=Z_{S_1\Delta S_2}$ because $Z\cdot Z=\mathds{1}$. The symmetric difference is associative. Thus $S_1\Delta S_2\Delta...\Delta S_n$ is well-defined and includes all elements that appear in an odd number of sets $S_i$ and we can write $Z_{S_1}Z_{S_2}...Z_{S_n} = Z_{S_1\Delta S_2\Delta...\Delta S_n}$.}. These expressions describe stabilizers of a new graph state with modified neighborhoods. To simplify the notation here and in the following sections, we assume for every stabilizer that $A, B$ are subtracted from all neighborhoods after the fusion as they have been destructively measured. In Eq.~\eqref{eq:xzzx_example_unconn_2}, for example, one would have to write $N(c_i)\setminus\{A,B\}$ instead of $N(c_i)$ otherwise.

An example of these graph transformation rules is given in Fig.~\ref{fig:intro}(a). To understand the corresponding graph transformation, note that qubits $2, 3$ are in the neighborhood of $A$ ($N(A)$) and thus Eq.~\eqref{eq:xzzx_example_unconn_0} applies. Therefore, their new neighborhood is the symmetric difference between their old neighborhood (minus qubit $A$) and the neighborhood of qubit $B$. In turn, the neighborhoods of the qubits $6, 7$ are transformed using Eq.~\eqref{eq:xzzx_example_unconn_1}. Except for the measured fusion qubits $A, B$, all other qubits are not in $N(A)$ or $N(B)$ and thus fulfill the condition in Eq.~\eqref{eq:xzzx_example_unconn_3} so that their neighborhood remains unchanged.

The fusion-induced transformation of a qubit's neighborhood can be illustrated by Venn diagrams. Fig.~\ref{fig:intro}(b) and Fig.~\ref{fig:Venn_main}(a) show Venn diagrams for the parity measurement $X_AZ_B\land Z_AX_B$. In Fig.~\ref{fig:intro}(b), the graph transformation is illustrated as an example for a qubit $a_1$ that is in the neighborhood $N(A)$ of the fusion qubit $A$, but not in the neighborhood $N(B)$ of the fusion qubit $B$. According to Eq.~\eqref{eq:xzzx_example_unconn_0}, the new neighborhood of $a_1$ is given by the symmetric difference $N(a_1)\Delta N(B)$ which is illustrated by the green area in the shown Venn diagram. Fig.~\ref{fig:Venn_main}(a) shows a Venn diagram for a qubit $c_i$ that is in the neighborhood of both fusion qubits ($c_i\in N(A)\cap N(B)$). In this case, Eq.~\eqref{eq:xzzx_example_unconn_2} applies and the neighborhood of qubit $c_i$ after the fusion is thus the symmetric difference $N(A)\Delta N(B)\Delta N(c_i)$ as illustrated by the green area in the shown Venn diagram.

\subsection{$X_AX_B\land Z_AZ_B$ ($A$, $B$ detached)}
\label{sec:XXZZ_detached}
In the following, we consider the widely-used fusion measuring the parities $X_AX_B\land Z_AZ_B$ upon success, corresponding to a standard linear-optical Bell state analyzer~\cite{Bartolucci2021, Paesani2022, Meng2023, Thomas2024}. We start with the case where both fusion qubits are detached and assume that all stabilizers in \cref{eq_stabs_all_1,eq_stabs_all_2,eq_stabs_all_3} exist, meaning that there is at least one qubit connected to the fusion qubit $A$ and not to $B$ ($N(A) \setminus N(B) \neq \emptyset$). The measured parity $X_AX_B$ anti-commutes with the stabilizers in \cref{eq_stabs_all_3,eq_stabs_all_4}. To make them commute we pick the stabilizer associated with a special neighbor $a_*$ that is connected to $A$ but not $B$, i.e. $a_*\in N(A)\setminus N(B)$ and multiply it on the other stabilizers in \cref{eq_stabs_all_3,eq_stabs_all_4}\footnote{Note that choosing a different special neighbor leads to a different but locally equivalent graph state. This illustrates the fact that there is generally some freedom of choice in the graph transformation rules: the same stabilizer states can be represented by different locally equivalent graphs~\cite{Adcock2020} with correspondingly different local Clifford gates.}. Next, $Z_AZ_B$ anti-commutes with the stabilizers in \cref{eq_stabs_all_1,eq_stabs_all_2} and thus we multiply Eq.~\eqref{eq_stabs_all_2} with Eq.~\eqref{eq_stabs_all_1}. After the two multiplications, the stabilizers are transformed into
\begin{flalign}
    &Z_{a_*}Z_{N(B) \Delta N(A)\setminus a_*}\label{eq:xxzz_s1}\\
    &\forall q_i \in \left(N(A)\Delta N(B)\right)\setminus a_*: 
    \notag\\
    &\begin{cases} q_i \notin N(a_*):X_{q_i}X_{a_*}Z_{N(q_i) \Delta N(a_*)}\\
    q_i \in N(a_*):Y_{q_i}Y_{a_*}Z_{N(q_i)\setminus a_* \Delta N(a_*)\setminus q_i}\label{eq:xxzz_unconn_raw_0}
    \end{cases}\\
    &\forall c_i \notin N(A)\Delta N(B): \notag\\
    &\begin{cases}c_i \notin N(a_*): X_{c_i}Z_{N(c_i)} \\
    c_i \in N(a_*):X_{c_i}Z_{a_*}Z_{N(c_i)\setminus a_*}\label{eq:xxzz_unconn_raw_1}
    \end{cases}
\end{flalign}
where we have dropped the Pauli matrices acting on the fusion qubits $A, B$ and explicitly written the operators acting on $a_*$. Furthermore, we have replaced the labels $a_i, b_i$ from~\cref{eq_stabs_all_3,eq_stabs_all_4} with a generic label $q_i$ in Eq.~\eqref{eq:xxzz_unconn_raw_0} since this qubit can be in either the neighborhood of the fusion qubit $A$ or the fusion qubit $B$.

The new stabilizers do not represent a graph state. To obtain a local Clifford equivalent graph state, we proceed by applying $H$ on qubit $a_*$. Multiplying the stabilizer in Eq.~\eqref{eq:xxzz_s1} on the stabilizers $q_i \in N(a_*)$ in Eq.~\eqref{eq:xxzz_unconn_raw_0}, and $c_i \in N(a_*)$ in Eq.~\eqref{eq:xxzz_unconn_raw_1}, we get the stabilizer generators of a new transformed graph state:
\begin{flalign}
    &X_{a_*} Z_{N(B) \Delta N(A)\setminus a_*}\label{eq_xxzz_unconn_a_st}\\
    &\forall q_i \in \left(N(A)\Delta N(B)\right)\setminus a_*:\notag\\
    &\begin{cases} q_i \notin N(a_*):X_{q_i}Z_{a_*}Z_{N(q_i)\Delta N(a_*)}\\
    q_i \in N(a_*):X_{q_i}Z_{a_*}Z_{N(q_i)\Delta N(A)\Delta N(a_*)\Delta N(B)} \label{eq_xxzz_unconn_0}
    \end{cases}\\
    &\forall c_i \notin N(A)\Delta N(B): \notag\\
    &\begin{cases}c_i \notin N(a_*): X_{c_i}Z_{N(c_i)} \\
    c_i \in N(a_*):X_{c_i}Z_{N(c_i)\Delta N(A)\Delta N(B)}
    \end{cases}\label{eq_xxzz_unconn_2}\\\notag
\end{flalign} 

This graph transformation rule only requires $N(A) \setminus N(B) \neq \emptyset$ (such that a special neighbor $a_*$ can be chosen) and thus applies to most cases. Since the measured parities are identical when interchanging the labels $A, B$, the case $N(A) \setminus N(B) = \emptyset \land N(B) \setminus N(A) \neq \emptyset$ is covered by the above graph transformation after swapping the labels $A, B$. In the final case when $N(B)\setminus N(A)= N(A)\setminus N(B) =\emptyset \land N(B)\cap N(A) \neq \emptyset$ ($A,B$ share all their neighbors), the graph state is simply updated by removing the fusion qubits since all stabilizers, except $S_A$ and $S_B$, commute with the measured parities\footnote{Note that we generally do not explicitly mention the cases where the two fusion qubits $A, B$ are not connected to any other qubit as, in this case, there is no graph transformation except removing these qubits from the graph.}. This is a particular case where the fusion success and the fusion failure coincide (see Appendix~\ref{sec:probability}).

As before, the graph transformations in \cref{eq_xxzz_unconn_a_st,eq_xxzz_unconn_0,eq_xxzz_unconn_2} can be represented by Venn diagrams. The case $c_i \in N(a_*)$ in Eq.~\eqref{eq_xxzz_unconn_2} is represented by the Venn diagram in Fig.~\ref{fig:Venn_main}(a) where the symmetric difference between the neighborhoods of three qubits in the original graph yields the neighborhood of the qubit after the fusion. Fig.~\ref{fig:Venn_main}(b) represents Eq.~\eqref{eq_xxzz_unconn_a_st} and Fig.~\ref{fig:Venn_main}(c) represents the new neighborhood of a qubit $q_i$ connected to one of the fusion qubits $A,B$ but not connected to the special neighbor qubit $a_*$ ($q_i \notin N(a_*)$ in Eq.~\eqref{eq_xxzz_unconn_0}). Finally, Fig.~\ref{fig:Venn_main}(d) represents the new neighborhood when the qubit is connected to the special neighbor ($q_i \in N(a_*)$ from Eq.~\eqref{eq_xxzz_unconn_0}).

An example of applying the corresponding graph transformation rule is given in Fig.~\ref{fig:intro}(a) where we have chosen the qubit number $2$ as the special neighbor ($a_*$). The Venn diagram in Fig.~\ref{fig:Venn_main}(a), resp. Eq.~\eqref{eq_xxzz_unconn_a_st} gives the new neighborhood of this qubit: the symmetric difference between the neighborhoods of both fusion qubits minus $a_*$ itself ($N(B) \Delta N(A)\setminus a_*$). The neighborhood of qubit $3$ is transformed according to the Venn diagram in Fig.~\ref{fig:Venn_main}(c), resp. Eq.~\eqref{eq_xxzz_unconn_0} with qubit $3 \notin N(a_*)$. Therefore, the new neighborhood of this qubit is the special neighbor qubit ($2$, resp. $a_*$) plus the symmetric difference between its old neighborhood and the neighborhood of qubit $2$. The symmetric difference contains qubit $4$ (part of the original neighborhood of qubit $3$) and a new connection to qubit $1$ (part of the neighborhood of the special neighbor qubit $2$). The neighborhood of qubits $6, 7$ also transforms according to the Venn diagram in Fig.~\ref{fig:Venn_main}(c) since the condition $q_i \notin N(a_*)$ from Eq.~\eqref{eq_xxzz_unconn_0} applies in both cases. The neighborhood of all other qubits remains unchanged.

\section{Applications}
\label{sec:appl}
\begin{figure}[!t]
\includegraphics[width=1.0\columnwidth]{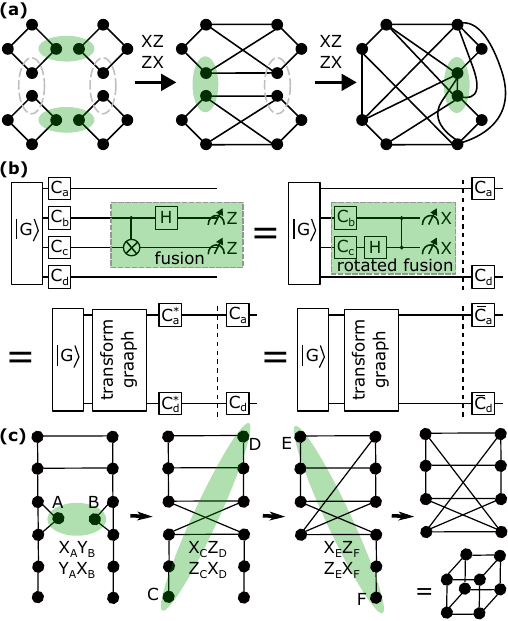}
\caption{\label{fig:fNet} Fusion networks and graph-state constructions. \textbf{(a)} Fusion network with four-qubit rings. If the first three fusions succeed (marked in green), the fusion qubits of the fourth fusion on the right are connected. \textbf{(b)} A successful fusion measuring $X_bX_c\land Z_bZ_c$ is applied to a stabilizer state (graph state $\ket{G}$ with local Clifford operations $C_i$ applied to it). The fusion (green box) is represented here by a circuit diagram including a CNOT gate, a $H$ gate, and $Z$-basis measurements (see Appendix~\ref{sec:circuit}). The local Clifford operations on the fusion qubits ($C_b, C_c$) change the fusion to a rotated fusion resulting in a different graph transformation (see Appendix~\ref{sec:setup}). For instance, a Hadamard gate on qubit $c$, $C_c=H_c$, would change the measured parities from $X_bX_c$ and $Z_bZ_c$ to $X_bH_c^{\dag}X_cH_c=X_bZ_c$ and $Z_bH_c^{\dag}Z_cH_c=Z_bX_c$. Since the other Clifford operations $C_a, C_d$ commute with the fusion circuit, they can be applied after the graph transformations (top right). Therefore, fusions applied to stabilizer states can be simulated by graph transformations (bottom part). Some fusion types can produce byproduct local Clifford gates (here $C_a^*, C_d^*$) and the local Clifford gates must be updated (here to $\Bar{C}_a, \Bar{C}_d$). \textbf{(c)} Generating the cube graph~\cite{Bell2022} using two different types of fusions. The assumed resource state can be generated deterministically with a single quantum emitter using the scheme from Refs.~\cite{Lindner2009,Tiurev2021} and local complementation~\cite{Cabello2011,Adcock2020} (see Appendix~\ref{sec:res_generate}).}
\end{figure}
In this section, we discuss some of the possible applications of the graph transformation rules derived in the previous section and Appendix~\ref{sec:fuse_unconnected},~\ref{sec:fuse_connected}: analyzing/simulating so-called fusion networks~\cite{Bartolucci2021, GimenoSegovia2015, Lobl2023} and constructing small graph states which can be used for error/loss-tolerant encoding in quantum communication~\cite{Azuma2015,Niu2022} or computing~\cite{Bell2022}.

\subsection{Analyzing fusion networks}
A fusion network describes a collection of small graph states (resource states) that are fused in a lattice topology that determines which qubits are fused with which~\cite{Bartolucci2021,GimenoSegovia2015,Lobl2023}. An example of a fusion network with four-qubit ring graphs as resource states and a square lattice topology is illustrated in Fig.~\ref{fig:fNet}(a). Graph transformation rules provide an intuitive way to understand the evolution of graph connectivity in fusion networks, which an analysis based on stabilizers~\cite{Aaronson2004,Fowler2009} does not directly provide.

To analyze a general fusion network, the graph transformation rules need to apply to arbitrary graphs. The reason is that even if fusions are performed between qubits of initially detached resource states, the fusion qubits may share part of their neighborhood or might be connected due to previous fusions as illustrated in Fig.~\ref{fig:fNet}(a).

Furthermore, it is necessary to have graph transformation rules for all cases in Table~\ref{tab:fusions} since a fusion causes a graph transformation plus potentially additional local Clifford gates on other qubits (see Section~\ref{sec:trafo}). These local Clifford gates can rotate future fusions, meaning that different parities are measured upon fusion success (see Appendix~\ref{sec:setup}). Fig.~\ref{fig:fNet}(b) illustrates this for two Clifford gates $C_b, C_c$. The local Clifford gates either rotate the fusion (like $C_b, C_c$) or act on other qubits, in which case they commute with the fusion (like $C_a, C_d$). The effect of the fusion can therefore be represented as a graph transformation followed by local Clifford operations (bottom part of Fig.~\ref{fig:fNet}(b)).

Finally, note that a graph state $\ket{G}$ with local Clifford gates $C_i$ applied to it (upper left in Fig.~\ref{fig:fNet}(b)) is a representation of an arbitrary stabilizer state~\cite{Gottesman1998, Hein2004, Adcock2020}. Therefore, graph transformations can be used to analyze the effect of fusions on any stabilizer state.

\subsection{Efficient simulation of fusion networks}
A stabilizer state can be represented as a graph state and local Clifford gates~\cite{Nest2004}. For this reason, a graph representation can be used to simulate Clifford circuits, which has been done in Refs.~\cite{Anders2006,Elliott2008,Elliott2009,Hu2022}. For most applications, the corresponding graphs are sparse, such that the average vertex degree $\Bar{d}$ of the graph is much smaller than the number of qubits $N$. The memory required to describe the graph is thus reduced to $\Bar{d}\cdot N$~\cite{Anders2006} compared to $\mathcal{O}(N^2)$ for representing stabilizers as a non-sparse matrix over $\mathds{F}_2$~\cite{Aaronson2004}. Furthermore, the time complexity for simulating gates and measurements reduces to $\mathcal{O}(\Bar{d}^2)$~\cite{Anders2006}. The Clifford circuit simulator from Ref.~\cite{Anders2006} is not explicitly made for fusions but could simulate the graph transformation upon a fusion by decomposing the fusion into corresponding gates and single-qubit measurements (see Appendix~\ref{sec:circuit}).

Our graph transformations provide the new graph neighborhoods after a fusion in an explicit form without decomposing it into a sequence of graph transformations. Compared to Ref.~\cite{Anders2006}, our graph transformations could therefore be more efficient for simulating fusion networks, although we expect the same scaling of the time complexity\footnote{Identical to the time complexity for gates and single-qubit measurements in  Ref.~\cite{Anders2006}, we also expect that the time complexity of our graph transformations is $\mathcal{O}(\Bar{d}^2)$ since, in such an operation, $\Bar{d}$ different qubits get a new neighborhood and this neighborhood is of size $\Bar{d}$.}. An issue is that fusion operations can be deterministic in some boundary cases (see Appendix~\ref{sec:deterministic_f}). In these cases, the fusion success probability is either one or zero depending on the sign of a stabilizer. Including stabilizer signs in our graph transformations would thus be relevant to building a full fusion network simulator.

\subsection{Constructing graph codes}
Small graph states representing graph codes can be used for all-photonic quantum communication~\cite{Azuma2015} as well as modular architectures for fault-tolerant photonic quantum computing~\cite{Bartolucci2021, Bell2022}. Certain graph states such as linear chains can be generated deterministically using a single quantum emitter with a spin~\cite{Lindner2009, Tiurev2021}. Recent experimental progress has been made~\cite{Thomas2022, Coste2023, Cogan2023, Meng2023b} but generating loss- and error-tolerant graph codes~\cite{Bell2022} is more challenging, as the deterministic generation of such states typically requires the ability to apply gates between quantum emitters~\cite{Li2022}. A promising alternative is to make small graph states deterministically with a quantum emitter~\cite{Lindner2009} and then create additional connections via fusions~\cite{Hilaire2023, Lee2023}.

Fig.~\ref{fig:fNet}(c) shows a corresponding construction of a cubic graph state that can be used for loss- and error-tolerant encoding~\cite{Bell2022}. Equipped with the transformation rules for various fusion types, we employ three fusions of two different types to construct the cube graph state from an input state that can be generated with a single quantum emitter and local gates (see Appendix~\ref{sec:res_generate}). Using different types of fusion may provide a degree of freedom that can be used to reduce the number of local gates or to adapt the construction to experimental boundary conditions. Without fusions, three quantum emitters and the ability to apply gates between them would be required (see Appendix~\ref{sec:res_generate}). In our construction, only three fusions are applied. This small number of fusions is advantageous because fusions are probabilistic, and fusion photons can suffer loss. More fusions mean more photons to start with and the loss of the additional fusion photons would diminish the loss- and error-tolerance of the graph code. Furthermore, the probabilistic nature of the fusions means that the overall chance of successfully generating the graph code is lowered by a factor of two with every additional fusion.

\section{Outlook/Summary}
We have derived graph transformation rules for graph states subject to all pairs of parity measurements from the Pauli group. These parity measurements correspond to the success case of probabilistic Bell state measurements (fusions). The derived graph transformations are relevant for the field of measurement-based~\cite{Raussendorf2001} and fusion-based~\cite{Bartolucci2021} quantum computing, which both rely on graph state transformations by measurements. The graph transformations are particularly useful for searching strategies for building dedicated graph states such as repeater graph states~\cite{Azuma2015} or graph codes~\cite{Bell2022} with minimal effort using fusion operations.

Parity measurements are not restricted to two qubits, and therefore a natural extension of our work is investigating parity measurements involving more qubits~\cite{Pankovich2023} (e.g. measuring three qubits and extracting three parities). Furthermore, similar work could be performed for different classes of quantum states. For the class of hypergraph states~\cite{Rossi2013}, for example, corresponding hypergraph transformation rules could potentially be developed.

\section{Acknowledgements}
We thank Yijian Meng, Luca Dellantonio, Andrew J. Jena, and Lane G. Gunderman for their feedback on parts of the manuscript. We are grateful for financial support from Danmarks Grundforskningsfond (DNRF 139, Hy-Q Center for Hybrid Quantum Networks) and Novo Nordisk Foundation (Challenge project “Solid-Q”). S.P. acknowledges funding from the Marie Skłodowska-Curie Fellowship project QSun (nr. 101063763), from the Villum Fonden research grant VIL50326, and support from the NNF Quantum Computing Programme.

\bibliography{main.bbl}

\makeatletter 
\renewcommand{\thefigure}{A\@arabic\c@figure}
\makeatother

\makeatletter 
\renewcommand{\thetable}{A\@arabic\c@table}
\makeatother

\makeatletter 
\renewcommand{\theequation}{A\@arabic\c@equation}
\makeatother

\onecolumngrid
\begin{appendices}
\count\footins = 1000

\section{Modified fusion setups}
\label{sec:setup}
Here we show that all parity measurements in Table~\ref{tab:fusions} can be performed by modifying the standard fusion setups in Figs.~\ref{fig:circiut}(a) with local Clifford gates $U_{A}, U_{B}$. The standard fusion setups measure $X_AX_B\land Z_AZ_B$ upon success. A gate $U_{A}$ transforms a state $\ket{G}$ into $U_{A}\ket{G}$ and thus its stabilizers $\{S_i\}$ are transformed as $S_i\rightarrow U_{A}S_iU_{A}^{\dag}$ (because $(U_{A}S_iU_{A}^{\dag})U_{A}\ket{G}=U_{A}S_i\ket{G}=U_{A}\ket{G}$). Rotating the measured parity as well as the state by applying a basis rotation $U_{A}^{\dag}$ to both yields the same final state up to this basis rotation and the same measurement result. Since we are not interested in the state of the measured qubit but only in how the rest of the graph transforms, we do not have to explicitly apply the opposite basis rotation after the fusion. The Bell state measurement with $U_{A}$ applied before can be interpreted in this way: keep the stabilizers the same ($U_{A}^{\dag}U_{A}S_iU_{A}^{\dag}U_{A}=S_i$) and rotate the Bell state measurement ($U_{A}^{\dag}X_AU_{A}\otimes X_B\land U_{A}^{\dag}Z_AU_{A}\otimes Z_B$)~\cite{Lobl2023}. Alternatively, one can consider the measured parities in the Heisenberg picture, where they are changed as $U_{A}^{\dag}X_AU_{A}\otimes X_B\land U_{A}^{\dag}Z_AU_{A}\otimes Z_B$, leaving the state unchanged.

By applying single-qubit Clifford gates from Table~\ref{tab:clifford-gates} the measurement pattern corresponding to the setup in Fig.~\ref{fig:circiut}(a) can be transformed into all the different parity measurements from Table~\ref{tab:fusions}. Possible gate choices are given in Table~\ref{tab:fusion-parities} together with the corresponding failure modes, in which case two single-qubit measurements are performed. Even if the parity measurement upon success is the same, the single-qubit measurements upon failure can differ, as one can see for the two rightmost cases in Table~\ref{tab:fusion-parities}. Furthermore, the choice of Clifford gates in Table~\ref{tab:clifford-gates} is not unique, and other choices can be found in Table 1 of Ref.~\cite{Hein2006}.

\begin{table}[h!]
\begin{center}
 \begin{tabular}{m{6em}ccccc}
 \hline
 \addlinespace[1.5ex]
  Gate & $H =\frac{X + Z}{\sqrt{2}} (\sqrt{Y}) $ &  $R = \sqrt{Z}$ & $Q = \sqrt{X}$ & $K = RH$ & $K^{\dagger} = HR^{\dagger}$  \\
  \addlinespace[1.5ex]
 \hline
 \hline
 \addlinespace[1.5ex]
  Matrix representation&
  $\frac{1}{\sqrt{2}}\begin{bmatrix}
      1 & 1 \\
      1 & - 1
  \end{bmatrix} 
  $ 
  & $\begin{bmatrix}
      1 & 0 \\
      0 & i
  \end{bmatrix} 
  $ & $\frac{1}{2}\begin{bmatrix}
      1-i & 1 +i \\
      1 + i & 1 - i
  \end{bmatrix} 
  $&  $\frac{1}{\sqrt{2}}\begin{bmatrix}
      1 & 1 \\
      i & -i
  \end{bmatrix} 
  $  & $\frac{1}{\sqrt{2}}\begin{bmatrix}
      1 & -i \\
      1 & i
  \end{bmatrix} 
  $
  \\
  \addlinespace[1.5ex]
 \hline
 \addlinespace[1.5ex]
 Pauli transformation & $X \leftrightarrow Z$ &  $X \leftrightarrow Y$ & $Y \leftrightarrow Z$ & $X \rightarrow Z \rightarrow Y \rightarrow X$ & $X \rightarrow Y \rightarrow Z \rightarrow X$ \\
 \addlinespace[1.5ex]
 \hline
 \end{tabular}
 \caption{\label{tab:clifford-gates}Local Clifford operators and their effect on the Pauli group. For the different operators, a matrix representation is shown, and which Pauli operators they interchange. Generally, the Pauli matrix $\sqrt{\sigma_i}$ interchanges the other two Pauli matrices while leaving $\sigma_i$ the same. For instance, $X, Z$ are interchanged by $\sqrt{Y}$ which could be chosen instead of the Hadamard gate $H$. A full version of this table including stabilizer signs can be found in Ref.~\cite{Hein2006}.}
 \end{center}
\end{table}

\begin{table}[h!]
\begin{center}
 \begin{tabular}{cccccccc}
 \hline
  Gates& $\mathds{1}_A\mathds{1}_B$ & $\mathds{1}_AH_B$ & $Q_AK_B$ & $R_AK_B^{\dag}$ & $Q_AR_B$ & $Q_AQ_B$ & $H_AH_B$  \\
  \addlinespace[1.5ex]
 \hline
 \addlinespace[1.5ex]
 Success & $X_AX_B$ & $X_AZ_B$ & $X_AY_B$ & $Y_AZ_B$ & $X_AY_B$ & $X_AX_B$ & $Z_AZ_B$  \\ 
  & $Z_AZ_B$ & $Z_AX_B$ & $Y_AX_B$ & $Z_AY_B$ & $Y_AZ_B$ &  $Y_AY_B$&  $X_AX_B$  \\
  \addlinespace[1.5ex]
 \hline
 \addlinespace[1.5ex]
 Failure & $Z_A\mathds{1}_B$ & $Z_A\mathds{1}_B$ & $Y_A\mathds{1}_B$ & $Z_A\mathds{1}_B$ & $Y_A\mathds{1}_B$ & $Y_A\mathds{1}_B$  & $X_A\mathds{1}_B$  \\ 
  & $\mathds{1}_AZ_B$ & $\mathds{1}_AX_B$ & $\mathds{1}_AX_B$ & $\mathds{1}_AY_B$ & $\mathds{1}_AZ_B$ & $\mathds{1}_AY_B$  & $\mathds{1}_AX_B$ \\
  \addlinespace[1.5ex]
 \hline
\end{tabular}
\caption{\label{tab:fusion-parities} Gates that need to be applied before the fusion circuit in Fig.~\ref{fig:circiut}(a) to implement the parity measurements in Table~\ref{tab:fusions}. The last two columns represent standard fusion upon success but have different failure modes.}
 \end{center}
\end{table}

\section{Circuit-based model of fusions}
\label{sec:circuit}
In the circuit-based quantum computing model, a fusion can be represented by local Clifford gates, a controlled-$Z$ gate, and single-qubit measurements. These operations can be represented as the following graph transformations: local graph complementation, adding additional local Clifford gates to the graph nodes, and adding links to the graph (controlled-$Z$ gate)~\cite{Hein2006,Anders2006} where the latter makes the fusion different from local operations and classical communication~\cite{Dahlberg2018}. A fusion can therefore be decomposed into such operations, which may also give an intuition for the graph transformations in some cases.

\begin{figure*}
\includegraphics[width=1.0\columnwidth]{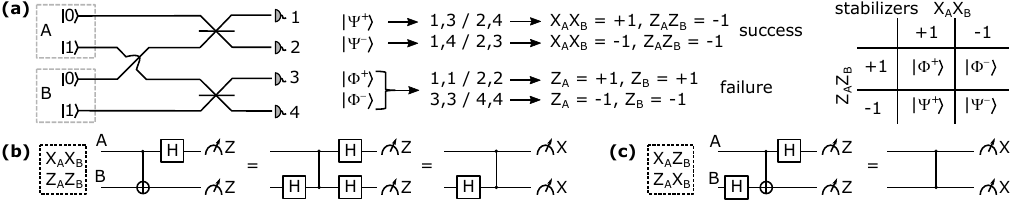}
\caption{\label{fig:circiut}\textbf{(a)} Standard setup of photonic fusion using dual-rail encoding (photon-number resolving detectors assumed). When detecting two photons at two different detectors, one projects in the subspace spanned by the two Bell states $\ket{\Psi^{\pm}}=\frac{1}{\sqrt{2}}(\ket{01}\pm\ket{10})$ as the other two Bell states $\ket{\Phi^{\pm}}=\frac{1}{\sqrt{2}}(\ket{00}\pm\ket{11})$ yield a different detection pattern. Therefore, $Z_AZ_B=-1$ is measured as this is a stabilizer of both states $\ket{\Psi^{\pm}}$ (see table on the right). Furthermore, $\ket{\Psi^{\pm}}$ can be distinguished by the measurement pattern leading to a measurement of $X_AX_B=+1$ when projecting on $\ket{\Psi^{+}}$ and $X_AX_B=-1$ when projecting on $\ket{\Psi^{-}}$. Since both stabilizers $X_AX_B$ and $Z_AZ_B$ are measured, these cases correspond to fusion success. When detecting two photons on the same detector, one projects into the subspace spanned by the other two Bell states $\ket{\Phi^{\pm}}=\frac{1}{\sqrt{2}}(\ket{00}\pm\ket{11})$ which measures $Z_AZ_B=+1$. A measurement of the stabilizer $X_AX_B$ is not obtained (fusion failure) as the two states $\ket{\Phi^{\pm}}$ yield the same detection pattern and thus cannot be distinguished. Instead, the detection pattern corresponds to two single-qubit measurements $Z_A\land Z_B$ with $Z_A=Z_B=\pm 1$. \textbf{(b)} Quantum circuit corresponding to a successful fusion for the setup in (a). \textbf{(c)} The parity measurements can be modified by single-qubit gates before the setup from (a). Applying $H$ in front of the setup in (a) corresponds, upon fusion success, to a controlled-$Z$ gate followed by two single-qubit $X$-basis measurements. This setup measures the parities $X_AZ_B\land Z_AX_B$.}
\end{figure*}

Figs.~\ref{fig:circiut}(b,c) show a representation of two Bell state measurements as a quantum circuit. A successful standard fusion is implemented by a CNOT-gate, a Hadamard, followed by two single-qubit measurements in the $Z$-basis, a configuration which measures the stabilizers $X_AX_B$ and $Z_AZ_B$ of the Bell state $\ket{\Phi_+}$. The circuit is equivalent to a Hadamard gate, followed by a controlled-$Z$ gate, and then two measurements in the $X$-basis (see Fig.~\ref{fig:circiut}(b)). Applying another Hadamard gate thus leads to a circuit with just a controlled-$Z$ gate followed by two single-qubit $X$-basis measurements of the fusion qubits (see Fig.~\ref{fig:circiut}(c))~\cite{Azuma2015}. This gives an intuitive interpretation of the parity measurement $X_AZ_B\land Z_AX_B$: draw a connection between the two unconnected fusion qubits, respectively remove an already existing connection (controlled-$Z$ gate), then do the graph transformations corresponding to the two $X$-basis measurements (see Ref.~\cite{Hein2006} and Appendix~\ref{sec:single_measure} for the graph transformation induced by single-qubit measurements). Note that the first $X$-basis measurement introduces a byproduct $H$ gate on one qubit, which can change the second $X$-basis measurement to a $Z$-basis measurement. Generally, expressing a fusion by concatenating graph transformations may give some intuition, yet does not explicitly provide the new neighborhood of every qubit. Our fusion rules from the main text and the Appendices~\ref{sec:fuse_unconnected},~\ref{sec:fuse_connected} do exactly that which is advantageous from a practical point of view.

\section{Fusion success probability}
\label{sec:probability}
In this section, we discuss the fusion success probability for unboosted type-II fusions~\cite{Browne2005, Gimeno2016}. We show that unboosted fusions succeed with probability $p_s=0.5$ in most cases, but that there are edge cases where success or failure is deterministic. A trivial example of such a deterministic fusion would be the case where the input state is a Bell state that is detached from the rest of the graph, e.g. if $\ket{\Psi^+}$ is sent to the standard fusion setup in Fig.~\ref{fig:circiut}(a), fusion deterministically succeeds and yields $Z_AZ_B=-1$ and $X_AX_B=+1$. However, we will show that there are a few other examples in which a deterministic outcome can occur.

Given a graph state $\ket{G}$ and two fusion qubits $A$ and $B$, the state $\ket{G}$ can be decomposed as\footnote{The relation follows by using twice that a graph state can be written as $\ket{G}=\frac{1}{\sqrt{2}}\left(\ket{0}_A\ket{\Omega}+\ket{1}_AZ^{N(A)}\ket{\Omega}\right)$, where $\ket{\Omega}$ is the graph state without qubit $A$.}:
\begin{align}
    \label{eq:DecomposedAB}
    \ket{G} = \frac{1}{2}(\ket{00}_{AB}\ket{\theta} + \ket{01}_{AB}Z_{N(B)\setminus A}\ket{\theta} +
    \ket{10}_{AB}Z_{N(A)\setminus B}\ket{\theta} \pm \ket{11}_{AB}Z_{(N(A)\setminus B) \Delta (N(B)\setminus A)}\ket{\theta}) \nonumber \\
    = \frac{1}{2}(\ket{00}_{AB}\ket{\theta^1} + \ket{01}_{AB}\ket{\theta^2} +
    \ket{10}_{AB}\ket{\theta^3} \pm \ket{11}_{AB}\ket{\theta^4})
\end{align}
where $\ket{\theta}$ is the induced subgraph of $\ket{G}$ with the qubits $A$ and $B$ removed, and $\ket{\theta^i}$ are associated graph basis states~\cite{Hein2006,Looi2008} with a $Z$-gate applied to some qubits ($\ket{\theta^1}=\ket{\theta}$, $\ket{\theta^2}=Z_{N(B)\setminus A}\ket{\theta}$, $\ket{\theta^3}=Z_{N(A)\setminus B}\ket{\theta}$, $\ket{\theta^4}=Z_{(N(A)\setminus B) \Delta (N(B)\setminus A)}\ket{\theta}$). The $\pm$ sign depends on whether $A$ and $B$ are connected ($-$) or not ($+$). A stabilizer state $\ket{S}$ is local-Clifford equivalent to at least one graph state $\ket{G}$~\cite{Hein2006} and we can thus use Eq.~\eqref{eq:DecomposedAB} to express a general stabilizer state as:
\begin{align}
    \label{eq:DecomposedAB_stab_st}
    \ket{S}=C^l\ket{G}=\frac{1}{2}(C^l_{AB}\ket{00}_{AB}\Tilde{C}^l\ket{\theta^1} + C^l_{AB}\ket{01}_{AB}\Tilde{C}^l\ket{\theta^2} + 
    C^l_{AB}\ket{10}_{AB}\Tilde{C}^l\ket{\theta^3} \pm C^l_{AB}\ket{11}_{AB}\Tilde{C}^l\ket{\theta^4})
\end{align}
with $C^l = C_1 \otimes C_2 \cdot \cdot \otimes C_N$ being a product of local Clifford gates on qubits $1$ to $N$, where $C^l_{AB}$ are the gates acting on the fusion qubits $A, B$ and $\Tilde{C}^l$ are the gates acting on other qubits. Note that in this section it will mostly be more convenient to think of the gates $C^l_{AB}$ rotating the states (Schrödinger picture) rather than the fusions as we have done in most of the manuscript.

In the graph $\ket{G}$, assume that the fusion qubits $A, B$ do not share their entire neighborhood ($N(A)\setminus B \neq N(B)\setminus A$) and are connected to more than just their fusion partner ($N(A)\setminus B \neq \emptyset \land N(B)\setminus A \neq \emptyset$). In this case, all states $\{\ket{\theta^{i}}\}$ are orthogonal graph basis states~\cite{Hein2006, Looi2008}. Then, the projection probability when projecting on $\ket{G}$ or $\ket{S}$ is $1/4$ for all Bell basis states, since the reduced density matrix obtained by tracing out all qubits except $A, B$ is $\rho_{A,B}=\frac{1}{4}\mathds{1}$. Therefore, the fusion success probability is $p_s=2\times\frac{1}{4}=0.5$ as only two of the Bell states can be distinguished by the Bell state analyzer (success) and two cannot (failure).

\subsection{Boundary case graph structures}
\label{sec:boundary_cases}
\begin{figure*}
\includegraphics[width=1.0\textwidth]{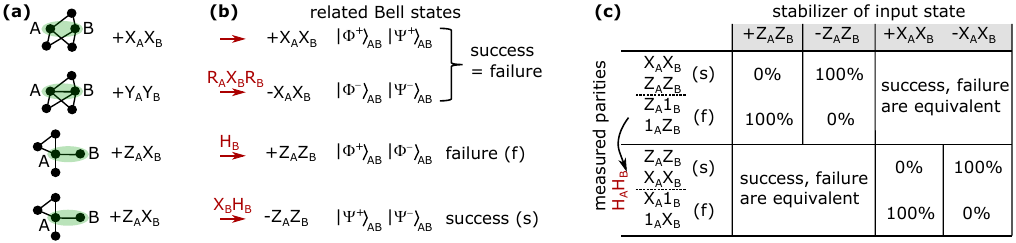}
\caption{\label{fig:p_fusion}\textbf{(a)} All graph boundary cases in which the fusions, illustrated by the green ellipses, can be deterministic (excluding isolated two-qubit states). In these cases, there is exactly one stabilizer $\Tilde{S}_{AB}$ with just support on $A, B$ that is shown on the right. \textbf{(b)} The stabilizer $\Tilde{S}_{AB}$ can be rotated by local Clifford gates to $\pm X_AX_B$ or $\pm Z_AZ_B$. When one of these operators is a stabilizer, the considered state $\ket{S}$ only contains two Bell states $\ket{b^{(1)}}_{AB}, \ket{b^{(2)}}_{AB}$ out of the four Bell states $\ket{\Psi^{\pm}}_{AB}=\frac{1}{\sqrt{2}}(\ket{01}\pm\ket{10})_{AB}$, $\ket{\Phi^{\pm}}_{AB}=\frac{1}{\sqrt{2}}(\ket{00}\pm\ket{11})_{AB}$, meaning $\braket{b^{(i)}_{AB}\mid S}= 0$ for the other Bell states. \textbf{(c)} Assuming a standard fusion setup (see Fig.~\ref{fig:circiut}(a)), deterministic fusion arises when $\pm Z_AZ_B$ is a stabilizer of the input state (upper part of the table). When $\pm X_AX_B$ is a stabilizer, fusion success and failure are equivalent as they result in the same state up to stabilizer signs. For a rotated fusion with $H_AH_B$ applied before the standard fusion setup, the roles of $\pm X_AX_B$ and $\pm Z_AZ_B$ are interchanged (lower part).}
\end{figure*}
There are cases where the states $\{\ket{\theta^{i}}\}$ in Eq.~\eqref{eq:DecomposedAB} do not form an orthonormal basis, and thus the previous derivation for the fusion probability $p_s=0.5$ does not apply. These cases correspond to fusion qubits with identical neighborhoods ($N(A)\setminus B = N(B)\setminus A$) or a fusion qubit that is at most connected to its fusion partner ($N(A)\setminus B= \emptyset$, or $N(B) \setminus A = \emptyset$). All these cases are illustrated in Fig.~\ref{fig:p_fusion}(a). In the following, we assume that only one of the above conditions is met, which means that the qubits $A, B$ are not a subgraph that is detached from the rest of the graph. In terms of stabilizers, this restriction means that there is only one stabilizer $\Tilde{S}_{AB}$ that only has support on $A,B$. To identify cases of deterministic fusion, we note that the following three statements are equivalent.
\begin{enumerate}
    \item The stabilizer state $\ket{S}$ is locally equivalent to a graph state fulfilling one or more of the following conditions: $N(A)\setminus B = N(B)\setminus A$, $N(A)\setminus B= \emptyset$, $N(B) \setminus A = \emptyset$
    \item In Eq.~\eqref{eq:DecomposedAB_stab_st}, some of the rotated graph basis states $\Tilde{C}^l\ket{\theta^i}$ coincide.
    \item $\ket{S}$ is stabilized by a stabilizer $\Tilde{S}_{AB}\neq\mathds{1}$ with support on only qubits $A, B$.
\end{enumerate}

\textit{Proof $-$} The first and second statements are equivalent since two graph basis states $Z_Q\ket{\theta}$ and $Z_P\ket{\theta}$ (with two subsets $P, Q$ of qubits from $\ket{\theta}$) are orthogonal, except if $Q=P$~\cite{Hein2006,Looi2008} as also mentioned in the previous section. Furthermore, the third statement follows from the first one: If $A, B$ are not isolated (have a neighbor), a graph state with $N(A)\setminus B=\emptyset$ has a stabilizer $X_AZ_B$, a graph state with $N(B)\setminus A=\emptyset$ has a stabilizer $Z_AX_B$, and a graph state with $N(A)\setminus B = N(B)\setminus A$ has a stabilizer $Y_AY_B$ or $X_AX_B$ depending on whether $A, B$ are connected or not. If either $A$ or $B$ are isolated, there is a stabilizer $X_A$ or $X_B$. Local gates do not change the fact that all these stabilizers have support on only $A, B$. Finally, assume that the third statement is true, so that there is at least one stabilizer $\Tilde{S}_{AB}\neq\mathds{1}$ with support only on $A, B$. Then some of the terms $\{\ket{\theta^{i}}\}$ in Eq.~\eqref{eq:DecomposedAB_stab_st} must coincide. The reason is that if all the states $\{\ket{\theta^{i}}\}$ were different (in which case they are orthogonal graph basis states), $\Tilde{S}_{AB}$ could only be a stabilizer if it stabilizes all basis states $C^l_{AB}\{\ket{00}_{AB}, \ket{01}_{AB}, \ket{10}_{AB}, \ket{11}_{AB}\}$ in which case $\Tilde{S}_{AB}=\mathds{1}$. Thus, the existence of a stabilizer $\Tilde{S}_{AB}\neq\mathds{1}$ with support on just $A, B$ implies that some of the states $\{\ket{\theta^{i}}\}$ are identical (second statement). Thus, all statements are equivalent\footnote{A stabilizer state can be locally Clifford equivalent to several graph states~\cite{Hein2004, Adcock2020}. If one of these graph states fulfills $N(A)\setminus B = N(B)\setminus A$, $N(A)\setminus B= \emptyset$, or $N(B) \setminus A = \emptyset$, this applies to all other locally Clifford equivalent graph states as well (although local complementations can change which of the three conditions applies). The reason is that a graph state stabilizer with support on just $A, B$ is equivalent to the graph state fulfilling at least one of the above conditions, and thus it would be a contradiction if a local operation (that does not change the support of the stabilizers) changes whether one of the three conditions is fulfilled.}. \qedsymbol{}

\subsection{Deterministic fusion}
\label{sec:deterministic_f}
From the previous sections, we know that the fusion is probabilistic with $p_s=0.5$ if there is no stabilizer with support on only the fusion qubits $A, B$. We are therefore interested in the case where there is one stabilizer $\Tilde{S}_{AB}$ with support on only $A,B$ implying that one of the conditions $N(A)\setminus B = N(B)\setminus A$, $N(A)\setminus B= \emptyset$, $N(B) \setminus A = \emptyset$ is fulfilled. We explain here under which condition the fusion is deterministic in such a case, either failing or succeeding with $p_s=0$ or $p_s=1$.

We assume the standard fusion setup (Fig.~\ref{fig:circiut}(a)) which measures the parities $X_AX_B, Z_AZ_B$ upon success (rotated fusions are discussed at the end of the section). The standard fusion can distinguish the Bell states $\ket{\Psi^{\pm}}$ by the detection pattern but cannot distinguish $\ket{\Phi^{\pm}}$. The states $\ket{\Phi^{\pm}}_{A,B}$ are stabilized by $+Z_AZ_B$ while $\ket{\Psi^{\pm}}_{A,B}$ are stabilized by $-Z_AZ_B$. $+Z_AZ_B$ being a stabilizer of the input state is therefore equivalent to only terms $\ket{\Phi^{\pm}}_{A,B}$ being present when expanding the state in the Bell basis over $A, B$ (Fig.~\ref{fig:p_fusion}(b)). In this and only in this case, the fusion fails deterministically ($p_s=0$) as the Bell state analyzer cannot distinguish $\ket{\Phi^{+}}_{A,B}$ from $\ket{\Phi^{-}}_{A,B}$ (Fig.~\ref{fig:p_fusion}(c)). In turn, $-Z_AZ_B$ being a stabilizer of the input state is equivalent to only terms of type $\ket{\Psi^{+}}_{A,B},\ket{\Psi^{-}}_{A,B}$ being present (which the standard Bell state analyzer can distinguish). In this and only in this case, the fusion is deterministically successful ($p_s=1$). The stabilizers $\pm Z_AZ_B$ thus herald deterministic fusion, which means that fusion deterministically fails (succeeds) if and only if $+Z_AZ_B$ ($-Z_AZ_B$) is a stabilizer (see Fig.~\ref{fig:p_fusion}(c)).

In the probabilistic fusion that is considered in most of this article, success and failure result in different graph transformation rules. As we shall now show, this is also the case for deterministic fusion. To this end, we consider two examples in which we will apply local gates to make $\pm Z_AZ_B$ a stabilizer and then apply the standard fusion (Fig.~\ref{fig:circiut}(a)) that upon success measures the parities $X_AX_B\land Z_AZ_B$.

First, assume a graph state where the fusion qubits $A$ and $B$ are not connected and share their entire neighborhood ($N(A) \setminus N(B) = N(B)\setminus N(A)$). We call the original graph state $\ket{G}$ and apply the gate $H_AZ_A \otimes H_B$ to it. (The corresponding rotated fusion measures $Z_AZ_B\land -X_AX_B$ on the original graph state, with fusion success being deterministic for a stabilizer $+X_AX_B$.) In the case considered, Eq.~\eqref{eq:DecomposedAB} becomes:
\begin{align}
    &H_AZ_A \otimes H_B\ket{G}= H_AZ_A \otimes H_B\frac{1}{\sqrt{2}}(\ket{\Phi^+}_{AB}\ket{\theta} + \ket{\Psi^{+}}_{AB}Z_{N(B)}\ket{\theta}) = \notag \\&\frac{1}{\sqrt{2}}(\ket{\Psi^{+}}_{AB}\ket{\theta} + \ket{\Psi^{-}}_{AB}Z_{N(B)}\ket{\theta})
\end{align}
This state only contains the two Bell states $\ket{\Psi^{\pm}}_{AB}$ and $-Z_AZ_B$ is therefore a stabilizer of the state. The standard fusion setup (see Fig.~\ref{fig:circiut}(a)) can distinguish both states, and fusing qubits $A, B$ will thus succeed with $p_s=1$. After the fusion, one obtains the states $\ket{\theta}$ or $Z_{N(B)}\ket{\theta}$, depending on whether one projects on $\ket{\Psi^+}_{AB}$ or $\ket{\Psi^-}_{AB}$, respectively. The graph transformation corresponds to removing $A,B$ from the original graph state $\ket{G}$ by two $Z$-basis measurements.

Now we instead apply $H_A \otimes H_B$ such that the corresponding rotated fusion measures $Z_AZ_B\land X_AX_B$ on the original graph state, with fusion failure being deterministic for a stabilizer $+X_AX_B$. Equivalently, rotating the graph state by the gate $H_A \otimes H_B$ gives:
\begin{align}
    \frac{1}{\sqrt{2}}(\ket{\Phi^{+}}_{AB}\ket{\theta} + \ket{\Phi^{-}}_{AB}Z_{N(B)}\ket{\theta})
\end{align}
This state only contains the two Bell states $\ket{\Phi^{\pm}}_{AB}$ and $+Z_AZ_B$ is therefore a stabilizer of the state. Since the standard fusion setup cannot distinguish the Bell states $\ket{\Phi^{\pm}}_{AB}$ the fusion deterministically fails ($p_s=0$). This corresponds to measuring $Z_A$ and $Z_B$ yielding the state $\ket{\theta} - Z_{N(B)}\ket{\theta}$ or $\ket{\theta} + Z_{N(B)}\ket{\theta}$ which would also be obtained by measuring $X_A$ and $X_B$ on the original graph state $\ket{G}$. (Note that in this boundary case, it would also be sufficient to measure just one of the qubits of $\ket{G}$ in the $X$-basis and the other one in any basis because the first measurement detaches the second fusion qubit from the rest of the graph.) Importantly, the corresponding graph transformation differs from the previous case where $-Z_AZ_B$ is a stabilizer, showing that the graph transformations for deterministic success and deterministic failure can differ.

So far we have assumed the standard fusion that, upon success, measures the parities $X_AX_B\land Z_AZ_B$. For rotated fusions, the stabilizers that herald deterministic fusion success and failure are correspondingly rotated. When applying, for instance, $H_AH_B$ before the fusion setup as above, then a stabilizer $+X_AX_B$ would herald deterministic failure and $-X_AX_B$ deterministic success (see the lower part of Fig.~\ref{fig:p_fusion}(c)). Instead, if the fusion is rotated with only $H_A$, then $+X_AZ_B$ heralds deterministic failure and $-X_AZ_B$ deterministic success.

Finally, assume that one of the fusion qubits $A, B$ is isolated but $A, B$ are not an isolated graph component (say qubit $A$ is isolated, $B$ is connected to some other qubits). Then the fusion will always be probabilistic. The reason is the following: since $A$ is detached, there is a stabilizer $S_A$ with only support on $A$. Assume that the fusion was deterministic and therefore there would be another stabilizer $\pm Z_AZ_B$. The two stabilizers $\pm Z_AZ_B$ and $S_A$ would have full rank on $A, B$, and so $A, B$ would be detached from the rest of the graph, which contradicts the initial assumption that they are not.

\subsection{Different stabilizers with support on only the fusion qubits}
\label{sec:different_stab}
In this section, we first argue that a stabilizer with support on only the fusion qubits implies that fusion success corresponds to two single-qubit measurements. Then, we show that if not $\pm Z_AZ_B$ but the other parity measured by a successful fusion, namely $\pm X_AX_B$, is a stabilizer, fusion success and failure give the same graph state (again assuming the standard fusion setup from Fig.~\ref{fig:circiut}(a)). In the corresponding cases, the outcome of the fusion is probabilistic, since Bell states of types $\ket{\Phi^{\pm}}_{A, B}$ and $ \ket{\Psi^{\pm}}_{A, B}$ are present. However, since success and failure give the same graph, it can be interpreted as deterministic.

The reader may have recognized that all the derived graph transformations for successful fusion correspond to single-qubit measurements for the boundary cases in Fig.~\ref{fig:p_fusion}(a) where there is a stabilizer $\Tilde{S}_{AB}$ with just support on $A, B$. In the following, we explain why this is the case: in the first boundary graph structure, $N(A)\setminus B = N(B)\setminus A$, Eq.~\eqref{eq:DecomposedAB_stab_st} becomes
\begin{equation}
\label{eq_example_xx_stab}
\frac{1}{\sqrt{2}}C^l_{AB}\Tilde{C}^l\left(\ket{\Phi^{\pm}}_{AB}\ket{\theta} + \ket{\Psi^+}_{AB} \ket{\Tilde{\theta}}\right),
\end{equation}
where the $\pm$ sign depends on whether $A, B$ are connected ($\ket{\Phi^{-}}$) or not ($\ket{\Phi^{+}}$) and $\ket{\Tilde{\theta}} = Z_{N(A)\setminus B}\ket{\theta}$. In the second possible case, $N(B) \setminus A = \emptyset$ (the case $N(A)\setminus B= \emptyset$ is analogous), Eq.~\eqref{eq:DecomposedAB_stab_st} becomes:
\begin{equation}
\label{eq_example_AB_stab}
\frac{1}{\sqrt{2}}C^l_{AB}\Tilde{C}^l\left(\ket{0}_A\ket{+}_B\ket{\theta} + \ket{1}_A\ket{-}_B\ket{\Tilde{\theta}}\right)
\end{equation}
where we have assumed that $A, B$ are connected as qubit $B$ would be isolated otherwise. In both cases, projecting on $\ket{\Psi^+}$ or $\ket{\Psi^-}$ upon fusion success can, up to a global phase, only yield stabilizer states of the form $\ket{\theta}, \ket{\Tilde{\theta}}, \frac{1}{\sqrt{2}}(\ket{\theta}+i^k \ket{\Tilde{\theta}})$ with $k\in\mathbb{N}_0$ (we derive this later in this paragraph). All these states also can be obtained from the corresponding graph state (with $C^l_{AB}=\mathds{1}$, $\Tilde{C}^l=\mathds{1}$) by measuring qubit $B$ in the $Z$-basis and doing another Pauli-basis measurement on qubit $A$\footnote{If single-qubit measurements in arbitrary bases were allowed, one could get any state of the form $\frac{1}{\sqrt{2}}(\ket{\theta}+e^{i\phi} \ket{\Tilde{\theta}})$ by single-qubit measurements since the two qubits $A, B$ are in a two-dimensional subspace for the corresponding boundary cases. However, we are restricted to measurements from the Pauli group here.}. When $C^l_{AB}\neq\mathds{1}$, these single-qubit Pauli measurements need to be rotated correspondingly. This shows that fusion success corresponds to single-qubit measurements in the mentioned boundary cases. We now prove the assumption that fusion success can only result in the mentioned states $\ket{\theta}, \ket{\Tilde{\theta}}, \frac{1}{\sqrt{2}}(\ket{\theta}+i^k \ket{\Tilde{\theta}})$: we first note that the inner product between two stabilizer states can only have absolute values of $0, (1/\sqrt{2})^s$ with $s\in\mathbb{N}_0$ being smaller or equal to the number of qubits~\cite{Aaronson2004, Garcia2012}. For a two-qubit system, the absolute value of projections such as $|\braket{\Psi^+\mid C^l_{AB}\Psi^+}|$ can therefore only be $0, 1/\sqrt{2}, 1/2, 1$ which restricts the possible states when projecting Eqs.~\eqref{eq_example_xx_stab},~\eqref{eq_example_AB_stab} on $\bra{\Psi^{\pm}}$ upon fusion success to a finite number of states\footnote{This also implies that the fusion success probability can only have discrete values for the mentioned boundary cases. By evaluating the sum of the projection probabilities of equations such as Eq.~\eqref{eq_example_xx_stab},~\eqref{eq_example_AB_stab} on $\ket{\Psi^+}$ and $\ket{\Psi^-}$, one finds that only $0, 0.5, 1$ are possible fusion success probabilities.}. Furthermore, post-fusion states where $\ket{\theta}$ and $\ket{\Tilde{\theta}}$ have prefactors of different nonzero amplitudes (e.g. $\frac{1}{\sqrt{3}}\ket{\theta}\pm \sqrt{\frac{2}{3}} Z_{N(A)\setminus B}\ket{\theta}$) are not possible because such states cannot be stabilizer states: projecting them on one of the orthogonal stabilizer states $\ket{\theta}$, $\ket{\Tilde{\theta}}$ would yield an inner product that is neither $0$ nor $(1/\sqrt{2})^s$. Finally, states of the form $\frac{1}{\sqrt{2}}(\ket{\theta}+e^{i\phi} \ket{\Tilde{\theta}})$ with $\phi\neq z\cdot\pi/2, z\in \mathbb{Z}$ cannot be stabilizer states either: projecting all qubits, except one in $N(A)\setminus B$, on $\ket{0}$ by Pauli $Z$-measurements would yield $\frac{1}{\sqrt{2}}(\ket{+}+e^{i\phi} \ket{-})$. If $\phi\neq z\cdot\pi/2, z\in \mathbb{Z}$, this is not an eigenstate of the Pauli operators $X, Y, Z$ and thus not a stabilizer state. Because only Pauli measurements were used, the original state cannot be a stabilizer state.

Now, we consider the case where the stabilizer with support on the fusion qubits is $\pm X_AX_B$, the other stabilizer measured by the standard fusion upon success. Then, the state contains only two Bell states: either $\ket{\Phi^{+}}_{A,B}, \ket{\Psi^{+}}_{A,B}$ (for a stabilizer $+X_AX_B$) or $\ket{\Phi^{-}}_{A,B}, \ket{\Psi^{-}}_{A,B}$ (for a stabilizer $-X_AX_B$) as illustrated in Fig.~\ref{fig:p_fusion}(b). In the case of fusion success, the parities $X_AX_B\land Z_AZ_B$ are measured by the standard fusion but since $\pm X_AX_B$ is a stabilizer, a measurement of the first parity $X_AX_B$ does nothing to the state. Measuring the second parity $Z_AZ_B$ yields a state where just one Bell state is left ($\ket{\Phi^{\pm}}_{A,B}\otimes\ket{S_1}$ or $\ket{\Psi^{\pm}}_{A,B}\otimes\ket{S_2}$). The fusion destructively measures the fusion qubits $A, B$ and the final state is thus $\ket{S_1}$ or $\ket{S_2}$ with no support on $A, B$. In the failure case of the standard fusion, $Z_A$ and $Z_B$ are measured, which is identical to measuring $Z_AZ_B$ and $Z_B$. However, measuring $Z_AZ_B$ gives again $\ket{\Phi^{\pm}}_{A,B}\otimes\ket{S_1}$ or $\ket{\Psi^{\pm}}_{A,B}\otimes\ket{S_2}$ as in the fusion success case. This is a product state between the qubits $A, B$ and the other qubits, and therefore it is irrelevant what other measurement (like $Z_B$ in the failure case or $X_AX_B$ in the success case) is performed on $A, B$ since the fusion destructively measures these qubits. Therefore, when $\pm X_AX_B$ is a stabilizer, one obtains the same states $\ket{S_1}$ or $\ket{S_2}$ for fusion success and fusion failure. Hence, assuming deterministic fusion success would give the correct graph transformation for the case of fusion failure. The graph transformation corresponds to removing the nodes $A, B$ from the graph by measuring $Z_A$ and $Z_B$.

Note that if the standard fusion is rotated by some local gate, the stabilizer for which fusion success and failure coincide will be $\pm X_AX_B$ rotated correspondingly. This is illustrated in Fig.~\ref{fig:p_fusion}(c) where for a fusion rotated by two gates $H_A, H_B$, the stabilizer $\pm Z_AZ_B$ becomes the one for which fusion success and failure coincide.

Finally, assume that the only stabilizer $\Tilde{S}_{AB}$ with support on only $A, B$ is not $\pm X_AX_B$. In this case, fusion success and failure can differ. An example is the case $C^l_{AB}=H_A$ in Eq.~\eqref{eq_example_xx_stab} giving the following state with stabilizer $+Z_AX_B$ (assuming unconnected fusion qubits, resp. $\ket{\Phi^{+}}$):
\begin{equation}
\frac{1}{2}\Tilde{C}^l\left((\ket{\Phi^{-}}+\ket{\Psi^{+}})_{AB}\ket{\theta} + (\ket{\Phi^{+}}+\ket{\Psi^{-}})_{AB} Z_{N(A)\setminus B}\ket{\theta}\right)
\end{equation}
In this case, fusion success has the effect of $Z$-basis measurements (node removal) on the associated graph state (state without the gate $C^l_{AB}=H_A$ gate). Fusion failure has the effect of measuring $X_A$ and $Z_B$ on the associated graph state. Since the $X$- and $Z$-basis measurements correspond to different graph transformations (see Ref.~\cite{Hein2006}, Appendix ~\ref{sec:single_measure}),  the graph states are generally different for success and failure even if there is a stabilizer $\Tilde{S}_{AB}$ with support on just $A, B$.

\subsection{Consequences of deterministic fusion}
\label{sec:determin_conseq}
We have seen that deterministic fusion can occur when $\pm Z_AZ_B$ is a stabilizer (for standard fusion). Deterministic fusion success occurs when $-Z_AZ_B$ is a stabilizer, deterministic failure occurs when $+Z_AZ_B$ is a stabilizer, and we found in the examples in Section~\ref{sec:deterministic_f} that the graph transformation upon deterministic success and deterministic failure can differ. For a precise simulation of arbitrary fusion networks where multiple graph states are fused, stabilizer signs should therefore be considered. Note, however, that there is no consequence for the validity of our graph transformations as a deterministic fusion failure case can be made a deterministic success by flipping a stabilizer sign with a local Pauli gate and our graph transformations do not specify this stabilizer sign. Thus, the derived graph transformations always correspond to possible operations, also in the boundary cases in Fig.~\ref{fig:p_fusion}(a). Applying single-qubit Pauli gates before a deterministic fusion could be a potential resource to make fusions deterministically successful in a fusion-based quantum computing architecture~\cite{Bartolucci2021, Paesani2022, Bombin2023b}. Furthermore, deterministic fusion only presents a boundary case, and applying the graph transformation rules without signs may thus provide a good approximation for the connectivity in a fusion network. In other words, $p_s=0.5$ typically applies on average for every single fusion, but conditioned on the outcome of other fusions, it might become deterministic resulting in correlations between the fusion outcomes~\cite{Bombin2023b}. The right strategy for dealing with a deterministic fusion depends on the exact physical situation, and in our implementation~\cite{Lobl2023} we therefore only give a warning in our source code~\cite{git2023} when a fusion can potentially be deterministic.

\section{Further graph transformations for unconnected fusion qubits}
\label{sec:fuse_unconnected}
\begin{figure*}[!t]
\includegraphics[width=1.0\columnwidth]{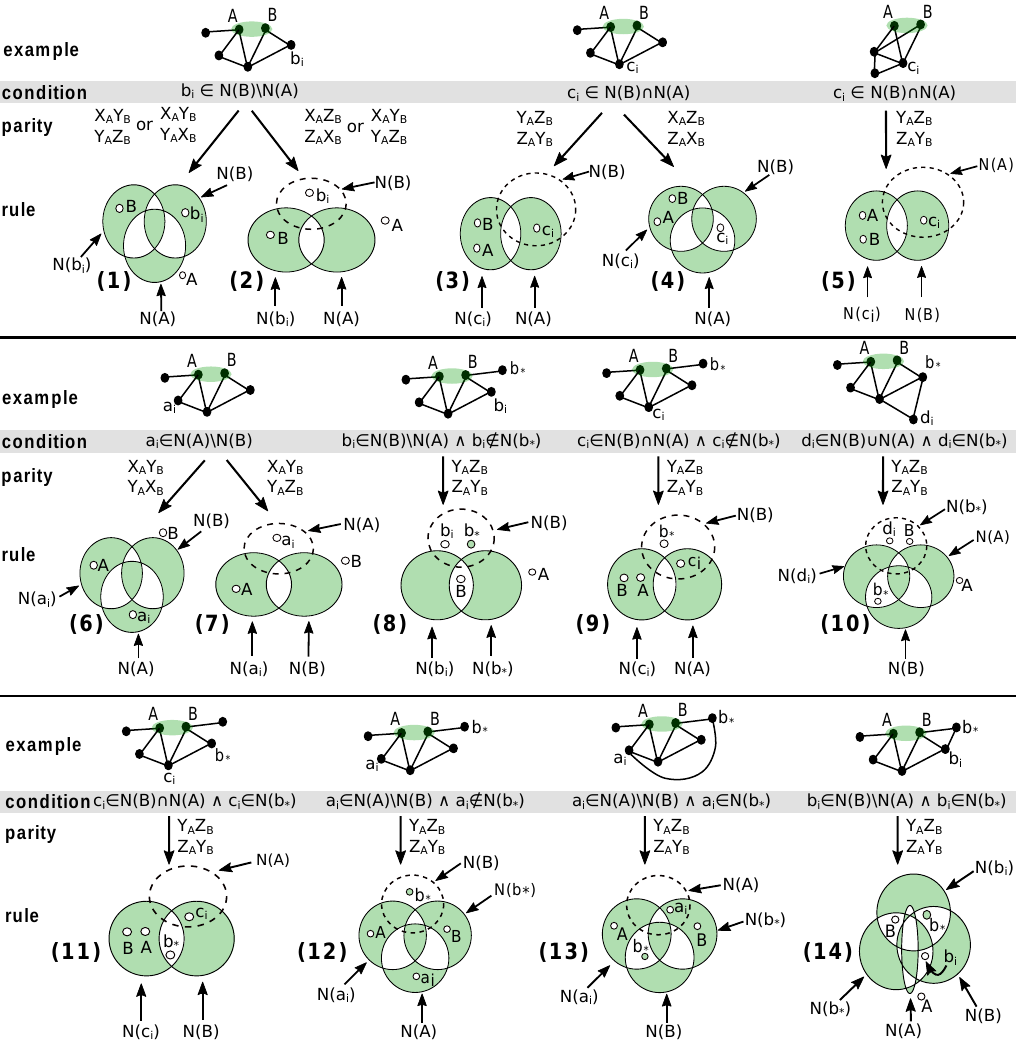}
\caption{\label{fig:appendix1}In subfigures (1)-(14) Venn diagrams for the remaining neighborhood transformations for the different fusion types with disconnected fusion qubits are illustrated. The subfigures represents the following equations: (1) \cref{eq_YXXY_D_in_symmetric_difference,eq_XYYZ_unconnected_2,eq_XYYZ_unconnected_1} (2) Eq.~\eqref{eq:xzzx_example_unconn_1}, (3) Eq.~\eqref{eq:XYYZ_unconnected_3}, (4) Eq.~\eqref{eq:xzzx_example_unconn_2}, (5) Eq.~\eqref{eq:YZZY_last_case_ci}, (6) Eq.~\eqref{eq_YXXY_D_in_symmetric_difference} for the case $q_i=a_i\in N(A)\setminus N(B)$, (7) Eq.~\eqref{eq:xzzx_example_unconn_0}, (8) Eq.~\eqref{eq:YZZY_D_bi_notin}, (9) Eq.~\eqref{eq:YZZY_D_ci_notin}, (10)  Eq.~\eqref{eq:YZZY_D_di_in}, (11) Eq.~\eqref{eq:YZZY_D_ci_in} , (12) Eq.~\eqref{eq:YZZY_D_ai_notin}, (13) Eq.~\eqref{eq:YZZY_D_ai_in}, (14) Eq.~\eqref{eq:YZZY_D_bi_in}.}
\end{figure*}
In the main text, we have derived some graph transformation rules for two fusion types from Section~\ref{sec:parity}. Here, we derive the graph transformations corresponding to the remaining cases. Fig.~\ref{fig:appendix1} shows the Venn diagrams that describe the corresponding neighborhood transformations. We do not consider the case $N(A)\setminus B = N(B)\setminus A = \emptyset$ (meaning $A$ and $B$ are disconnected from the rest of the graph)
since the measurement of the fusion qubits has no effect on the rest of the graph in this case.

\subsection{$X_AY_B\land Y_AX_B$ ($A$, $B$ detached)}
\label{sec:xyyx_detached}
We start by considering the case where all the stabilizers in \cref{eq_stabs_all_1,eq_stabs_all_2,eq_stabs_all_3,eq_stabs_all_4,eq_stabs_all_5} exist. The first measured parity, $X_AY_B$, anti-commutes with the stabilizers in \cref{eq_stabs_all_2,eq_stabs_all_3,eq_stabs_all_4} and therefore Eq.~\eqref{eq_stabs_all_2} is multiplied with the other two sets of stabilizers. The second parity, $X_AY_B$, anti-commutes with the stabilizers in \cref{eq_stabs_all_1,eq_stabs_all_3,eq_stabs_all_4}. Thus, we multiply Eq.~\eqref{eq_stabs_all_1} on \cref{eq_stabs_all_3,eq_stabs_all_4}. Disregarding the Pauli operators acting on the measured qubits $A, B$ yields the transformed stabilizers:
\begin{flalign}
    &\forall a_i \in N(A)\setminus N(B):Y_{a_i}Z_{N(a_i)}Z_{N(B)}Z_{N(A)\setminus a_i} \label{eq:YXXY-ai}\\
    &\forall b_i \in N(B)\setminus N(A):Y_{b_i}Z_{N(b_i)}Z_{N(A)}Z_{N(B)\setminus b_i} \label{eq:YXXY-bi}\\
    &\forall c_i \in N(A)\cap N(B):X_{c_i}Z_{N(c_i)}\label{eq:YXXY-ci}\\
    &\forall d_i \notin N(B)\cup N(A): X_{d_i}Z_{N(d_i)} \label{eq:YXXY-di}
\end{flalign}
Applying $R$ on all qubits $b_i$ and $a_i$ yields a graph state:
\begin{flalign}
    &\forall q_i \in N(A)\Delta N(B):X_{q_i}Z_{N(q_i)\Delta \left(N(A)\setminus q_i\right)\Delta\left(N(B)\setminus q_i\right)} \label{eq_YXXY_D_in_symmetric_difference}\\
    &\forall c_i \notin N(A)\Delta N(B):X_{c_i}Z_{N(c_i)}\label{eq_YXXY_D_not_in_symmetric_difference}
\end{flalign}
Note that only the stabilizers $S(A)$ and $S(B)$ are used to make anti-commuting stabilizers commute and thus the above update rule is valid for any neighborhood configuration.

\subsection{$X_AY_B\land Y_AZ_B$ ($A$, $B$ detached)}
\label{sec:xyyz_connect}
Multiplying $S_A$ from Eq.~\eqref{eq_stabs_all_1} on the stabilizers in \cref{eq_stabs_all_4,eq_stabs_all_5} and multiplying $S_B$ from Eq.~\eqref{eq_stabs_all_2} on the stabilizers in \cref{eq_stabs_all_3,eq_stabs_all_4} yields stabilizers that commute with both measured parities. After applying single-qubit $R$-gates (transforming Pauli $Y$ to Pauli $X$) to the qubits $b_i$ in $S_{b_i}$ and $c_i$ in $S_{c_i}$, these equations become the stabilizer generators of a new graph state:
\begin{align}
    &\forall a_i \in N(A)\setminus N(B): S_{a_i}=X_{a_i}Z_{N(a_i)\Delta N(B)}\label{eq_XYYZ_unconnected_1}\\
    &\forall b_i \in N(B)\setminus N(A): S_{b_i}=X_{b_i}Z_{N(b_i)\Delta (N(B)\setminus b_i)\Delta N(A)}\label{eq_XYYZ_unconnected_2}\\
    &\forall c_i \in N(A)\cap N(B): S_{c_i}=X_{c_i}Z_{N(c_i)\Delta \left(N(A)\setminus c_i\right)}\label{eq:XYYZ_unconnected_3}\\
    &\forall d_i \notin N(B)\cup N(A): S_{d_i}=X_{d_i}Z_{N(d_i)}
\end{align}
Here we have dropped $S_A, S_B$ and the operators on the measured qubits $A, B$ in the other stabilizers. The associated Venn diagrams are very similar to Figs.~\ref{fig:Venn_main}(a) and~\ref{fig:intro}(b).

\subsection{$Y_AZ_B\land Z_AY_B$ ($A$, $B$ detached)}
\subsubsection{$N(B) \setminus N(A) \neq \emptyset$}
\label{sec:yzzy_a}
We start by considering the case where at least all the stabilizers in \cref{eq_stabs_all_1,eq_stabs_all_2,eq_stabs_all_4}. Then
the first measured parity, $Y_AZ_B$, anti-commutes with the stabilizers in \cref{eq_stabs_all_1,eq_stabs_all_2,eq_stabs_all_3,eq_stabs_all_5} and thus we multiply the stabilizer $S_B$ in Eq.~\eqref{eq_stabs_all_2} on the other three sets of stabilizers. After the multiplication and dropping $S_B$ the stabilizers given in Eq.~\eqref{eq_stabs_all_3}, Eq.~\eqref{eq_stabs_all_4} anti-commute with $Z_AY_B$. Therefore, we pick one stabilizer in Eq.~\eqref{eq_stabs_all_4}, which we denote $S_{b_*}$, multiply it with the rest, and then drop it. Without the measured fusion qubits $A,B$, the stabilizers in \cref{eq_stabs_all_1,eq_stabs_all_3,eq_stabs_all_5,eq_stabs_all_6} are transformed as:
\begin{align}
    &Z_{N(A)\Delta N(B)} \label{eq:YZZY-B}\\
    &\forall a_i \in N(A) \setminus N(B):\notag\\
    &\hspace{0.5cm}\textbf{if } a_i\notin N(b_*): X_{a_i}Y_{b_*}Z_{N(a_i)\Delta (N(B) \setminus b_*)\Delta N(b_*)}\label{eq_ai-YZZY_detached}\\
    &\hspace{0.5cm}\textbf{else }: Y_{a_i}X_{b_*}Z_{N(a_i)\Delta N(B)\Delta (N(b_*)\setminus a_i)}\label{eq_aj-YZZY_detached}\\
    &\forall b_i \in N(B) \setminus N(A)\setminus b_*:\notag\\
    &\hspace{0.5cm}\textbf{if } b_i\notin N(b_*): X_{b_i}X_{b_*}Z_{N(b_i) \Delta N(b_*)}\\
    &\hspace{0.5cm}\textbf{else }: Y_{b_i}Y_{b_*}Z_{(N(b_i)\setminus b_*) \Delta (N(b_*)\setminus b_i)}\\
    &\forall c_i \in N(A) \cap N(B):\notag\\
    &\hspace{0.5cm}\textbf{if } c_i\notin N(b_*): Y_{c_i}Z_{b_*}Z_{N(c_i) \Delta (N(B) \setminus b_* \setminus c_i)}\label{eq_ci-YZZY_detached}\\
    &\hspace{0.5cm}\textbf{else }: Y_{c_i}Z_{N(c_i) \Delta (N(B)\setminus c_i)} \label{eq_cj-YZZY_detached}\\
    &\forall d_i \notin N(B)\cup N(A):\notag\\
    &\hspace{0.5cm}\textbf{if } d_i\notin N(b_*): X_{d_i}Z_{N(d_i)}\\
    &\hspace{0.5cm}\textbf{else }: X_{d_i}Z_{b_*}Z_{N(d_i)\setminus b_*}
\end{align}
We proceed by 1) applying $H$ on $b_*$, 2) multiplying Eq.~\eqref{eq:YZZY-B} to all stabilizers containing a $X_{b_*}$ or $Y_{b_*}$ (after applying $H$ on $b_*$), and 3) applying $R$ on the qubits $a_i$ and $c_i$ in \cref{eq_ai-YZZY_detached,eq_aj-YZZY_detached,eq_ci-YZZY_detached,eq_cj-YZZY_detached}. This yields the stabilizer generators representing the transformed graph state:
\begin{align}
    &X_{b_*}Z_{N(A)\Delta (N(B)\setminus b_*)} \label{eq:YZZY_D_b_special} \\
    &\forall a_i \in N(A) \setminus N(B):\notag\\
    &\hspace{0.5cm}\textbf{if } a_i\notin N(b_*): X_{a_i}Z_{b_*}Z_{N(a_i)\Delta (N(A)\setminus a_i)\Delta N(b_*)} \label{eq:YZZY_D_ai_notin}\\
    &\hspace{0.5cm}\textbf{else }: X_{a_i}Z_{b_*}Z_{N(a_i)\Delta N(B)\Delta (N(b_*)\setminus a_i)}\label{eq:YZZY_D_ai_in}\\
    &\forall b_i \in N(B) \setminus N(A)\setminus b*:\notag\\
    &\hspace{0.5cm}\textbf{if } b_i\notin N(b_*): X_{b_i}Z_{b_*}Z_{N(b_i) \Delta N(b_*)} \label{eq:YZZY_D_bi_notin}\\
    &\hspace{0.5cm}\textbf{else }: X_{b_i}Z_{b_*}Z_{N(b_i)\Delta N(A)\Delta N(B)\Delta N(b_*)} \label{eq:YZZY_D_bi_in}\\
    &\forall c_i \in N(A) \cap N(B):\notag\\
    &\hspace{0.5cm}\textbf{if } c_i\notin N(b_*):  X_{c_i}Z_{N(c_i) \Delta (N(A) \setminus c_i)} \label{eq:YZZY_D_ci_notin}\\
    &\hspace{0.5cm}\textbf{else }: X_{c_i}Z_{N(c_i) \Delta (N(B)\setminus c_i)} \label{eq:YZZY_D_ci_in} \\
    &\forall d_i \notin N(B)\cup N(A):\notag\\
    &\hspace{0.5cm}\textbf{if } d_i\notin N(b_*): X_{d_i}Z_{N(d_i)} \label{eq:YZZY_D_di_notin}\\
    &\hspace{0.5cm}\textbf{else }: X_{d_i}Z_{N(d_i)\Delta N(B)\Delta N(A)} \label{eq:YZZY_D_di_in}
\end{align}

\subsubsection{$N(B) \setminus N(A) =\emptyset\land N(A) \setminus N(B) \neq\emptyset$}
For $N(B) \setminus N(A) = \emptyset$ the special neighbor $b_*$ in the previous section has to be changed to $a_*$. Since the measured parities $Y_AZ_B\land Z_AY_B$ are symmetric concerning interchanging $A, B$, the relevant transformation rules can be obtained by swapping the labels $A, B$ in the previous section.

\subsubsection{$N(B) \setminus N(A)=\emptyset\land N(A) \setminus N(B)=\emptyset\land N(B)\cap N(A)\neq\emptyset$}
Here, $A$ and $B$ completely share their neighborhood, i.e. the stabilizers in \cref{eq_stabs_all_3,eq_stabs_all_4} do not exist, but the stabilizers in Eq.~\eqref{eq_stabs_all_5} do. However, the graph transformation is analogous to Section~\ref{sec:yzzy_a} as the derivation coincides (up to choosing a special neighbor $b_*$). As for the case above, the stabilizers in \cref{eq_stabs_all_1,eq_stabs_all_2,eq_stabs_all_5} anti-commutes with $Y_AZ_B$ and thus we multiply Eq.~\eqref{eq_stabs_all_2} on the other three sets of stabilizers. After the multiplication, all stabilizers commute with $Z_AY_B$. We apply $R$ on all qubits $c_i\in N(A)\cap N(B)$ to transform $Y_{c_i} \xrightarrow{} X_{c_i}$. Dropping the measured qubits $A, B$ yields graph state stabilizers: 
\begin{align}
    &\forall c_i \in N(A)\cap N(B): X_{c_i}Z_{N(c_i) \ \Delta (N(B)\setminus c_i)}\label{eq:YZZY_last_case_ci}\\
    &\forall d_i \notin N(B)\cup N(A): X_{d_i}Z_{N(d_i)}
\end{align}

\section{Graph transformations for connected fusion qubits}
\label{sec:fuse_connected}
In the main text and in Appendix~\ref{sec:fuse_unconnected}, we have derived graph transformation rules when both fusion qubits are detached. In this section, we derive the corresponding rules for connected fusion qubits. When the fusion qubits $A$ and $B$ of the graph state are connected, the different stabilizer generators read:
\begin{align}
    &S_A = X_A Z_B Z_{N(A)\setminus B} \label{eq_stabs_connected_1}\\
    &S_B = X_B Z_A Z_{N(B)\setminus A}\label{eq_stabs_connected_2}\\
    &\forall a_i \in N(A)\setminus B\setminus N(B): S_{a_i}=X_{a_i}Z_{N(a_i)\setminus A}Z_A\label{eq_stabs_connected_3}\\
    &\forall b_i \in N(B)\setminus A\setminus N(A): S_{b_i}=X_{b_i}Z_{N(b_i)\setminus B}Z_B\label{eq_stabs_connected_4}\\
    &\forall c_i \in N(A)\cap N(B): S_{c_i}=X_{c_i}Z_{N(c_i)\setminus A \setminus B}Z_AZ_B\label{eq_stabs_connected_5}\\
    &\forall d_i \notin N(B)\cup N(A): S_{d_i}=X_{d_i}Z_{N(d_i)}\label{eq_stabs_connected_6}
\end{align}

\subsection{$X_AZ_B\land Z_AX_B$ ($A$, $B$ connected)}
In the main text, we have considered the fusion type $X_AZ_B\land Z_AX_B$ for the case that the two fusion qubits are not connected. The case that the two fusion qubits are connected is more involved as Eq.~\eqref{eq_stabs_connected_1} and Eq.~\eqref{eq_stabs_connected_2} both commute with the measured parities and therefore multiplication with none of them can be used to make anticommuting stabilizers commute. Starting from \cref{eq_stabs_connected_1,eq_stabs_connected_2,eq_stabs_connected_3,eq_stabs_connected_4,eq_stabs_connected_5,eq_stabs_connected_6}, we derive the corresponding rules when the fusion qubits $A$, $B$ are connected. Depending on the neighborhood of the fusion qubits, not all these types of stabilizers may exist leading to different transformations of the stabilizers/graphs. These cases are considered in the following.

\subsubsection{$(N(B)\setminus A = \emptyset\land N(A)\setminus B \neq\emptyset)\lor(N(A)\setminus B = \emptyset\land N(B)\setminus A \neq\emptyset)$}
Assume that qubit $B$ is only connected to qubit $A$ and qubit $A$ has at least one additional neighbor ($N(B)\setminus A = \emptyset\land N(A)\setminus B \neq\emptyset$). The case $N(A)\setminus B = \emptyset\land N(B)\setminus A \neq\emptyset$ is analogous by interchanging the labels $A, B$. In this case, the stabilizer $S_B$ in Eq.~\eqref{eq_stabs_connected_2} is just one of the measured parities and can be removed. Furthermore, a stabilizer of type $S_{b_i}$ in Eq.~\eqref{eq_stabs_connected_4} and $S_{c_i}$ in Eq.~\eqref{eq_stabs_connected_5} does not exist. Looking through the list of stabilizers, $S_{a_i}$ in Eq.~\eqref{eq_stabs_connected_3} is the first one that anticommutes with the measured parity $X_AZ_B$. Therefore, we chose a special neighbor $a_*\in N(A)\setminus B\setminus N(B)$ and multiply $S_{a_*}$ on all other stabilizers that anticommute with $S_{a_*}$. The remaining stabilizers commute with the second measured parity $Z_AX_B$ and we drop the part of them that acts on the measured qubits $A, B$. This gives:
\begin{align}
    &Z_{N(A)}\label{eq_xzzx_b_isolated_1}\\
    &\forall a_i \in N(A)\setminus B\setminus N(B): X_{a_i}X_{a_*}Z_{N(a_i)\Delta N(a_*)}\label{eq_xzzx_b_isolated_2}\\
    &\forall d_i \notin N(B)\cup N(A): S_{d_i}=X_{d_i}Z_{N(d_i)\label{eq_xzzx_b_isolated_3}}
\end{align}
Writing the support on qubit $a_*$ explicitly and using $N(B)=A\notin N(A)$ yields the following:
\begin{align}
    &Z_{a_*}Z_{N(A)\setminus a_*}\label{eq:xzzx_case2}\\
    &\forall a_i \in N(A)\setminus B\setminus a_*:\notag\\
    &\hspace{0.5cm}\textbf{if } a_i\notin N(a_*): X_{a_i}X_{a_*}Z_{N(a_i)\Delta N(a_*)}\\
    &\hspace{0.5cm}\textbf{else }: Y_{a_i}Y_{a_*}Z_{(N(a_i)\setminus a_*)\Delta (N(a_*)\setminus a_i)}\\
    &\forall d_i \notin N(B)\cup N(A):\notag\\
    &\hspace{0.5cm}\textbf{if } d_i\notin N(a_*): X_{d_i}Z_{N(d_i)}\\
    &\hspace{0.5cm}\textbf{else }: X_{d_i}Z_{a_*}Z_{N(d_i)\setminus a_*}
\end{align}
Applying $H_{a_*}$, then multiplying Eq.~\eqref{eq:xzzx_case2} on all other stabilizers including $X_{a_*}$ or $Y_{a_*}$ gives:
\begin{align}
    &X_{a_*}Z_{N(A)\setminus a_*} \label{eq:66}\\
    &\forall a_i \in N(A)\setminus B\setminus a_*:\notag\\
    &\hspace{0.5cm}\textbf{if } a_i\notin N(a_*): X_{a_i}Z_{a_*}Z_{N(a_i)\Delta N(a_*)} \label{eq:67}\\
    &\hspace{0.5cm}\textbf{else }: X_{a_i}Z_{a_*}Z_{N(a_i), N(a_*), N(A)}\label{eq:68}\\
    &\forall d_i \notin N(B)\cup N(A):\notag\\
    &\hspace{0.5cm}\textbf{if } d_i\notin N(a_*): X_{d_i}Z_{N(d_i)}\\
    &\hspace{0.5cm}\textbf{else }: X_{d_i}Z_{N(d_i)\Delta N(A)}
\end{align}
which are the stabilizers of a graph state. In this particular case, the graph transformation is identical to measuring qubit $A$ in the $X$-basis (see Section~\ref{sec:single_measure}) which follows from the discussion in Section~\ref{sec:different_stab}. For simplicity, we do not explicitly draw Venn diagrams for the different cases of connected fusion qubits in the following. However, all graph transformations can still be directly translated into a Venn diagram picture as before.

\subsubsection{$N(A)\setminus B\setminus N(B)=N(B)\setminus A\setminus N(A)=\emptyset\land N(A)\cap N(B)\neq\emptyset$}
Here, qubits $A, B$ share their entire neighborhood (not considering $A, B$ themselves). Consequently, stabilizers of type $S_{a_i}$ from Eq.~\eqref{eq_stabs_connected_3} and $S_{b_i}$ from Eq.~\eqref{eq_stabs_connected_4} do not exist (and so no special neighbor $a_*$ can be selected as in the previous sub-section). Since the stabilizers in \cref{eq_stabs_connected_1,eq_stabs_connected_2} are identical except for the operators on the measured qubits $A, B$ and commute with both measured parities, we just keep Eq.~\eqref{eq_stabs_connected_1} and discard Eq.~\eqref{eq_stabs_connected_2} upon the fusion. We chose a special neighbor $c_*\in N(A)\cap N(B)$ and multiply $S_{c_*}$ on all other stabilizers of type $S_{c_i}$, making them commute with both measured parities. Dropping the operators on the measured qubits $A, B$ yields:
\begin{align}
    &Z_{N(A)}\\
    &\forall c_i \in N(A)\cap N(B): X_{c_i}X_{c_*}Z_{N(c_i)\Delta N(c_*)}\\
    &\forall d_i \notin N(B)\cup N(A): S_{d_i}=X_{d_i}Z_{N(d_i)}
\end{align}
These equations are very similar to \cref{eq_xzzx_b_isolated_1,eq_xzzx_b_isolated_2,eq_xzzx_b_isolated_3} from the previous section (with stabilizers of type $S_{c_i}$ instead of type $S_{a_i}$). After applying the gate $H_{c_*}$, the graph transformation rules are, therefore, identical with neighbors of type $c_i$ rather than type $a_i$:
\begin{align}
    &X_{c_*}Z_{N(A)\setminus c_*}\\
    &\forall c_i \in (N(A)\cap N(B))\setminus c_*:\notag\\
    &\hspace{0.5cm}\textbf{if } c_i\notin N(c_*): X_{c_i}Z_{c_*}Z_{N(c_i)\Delta N(c_*)} \label{eq:75}\\
    &\hspace{0.5cm}\textbf{else }: X_{c_i}Z_{c_*}Z_{N(c_i), N(c_*), N(A)}\\
    &\forall d_i \notin N(B)\cup N(A):\notag\\
    &\hspace{0.5cm}\textbf{if } d_i\notin N(c_*): X_{d_i}Z_{N(d_i)} \label{eq:xzzx_C_first_case_ci_in}\\
    &\hspace{0.5cm}\textbf{else }: X_{d_i}Z_{N(d_i)\Delta N(A)} \label{eq:xzzx_C_first_case_di_not_in}
\end{align}

\subsubsection{$N(A)\setminus B\setminus N(B)\neq\emptyset\land N(B)\setminus A\setminus N(A)\neq\emptyset$}
Here, we consider the most typical case that both fusion qubits have at least one neighbor that is not connected to the other fusion qubit, and thus stabilizers of type Eq.~\eqref{eq_stabs_connected_3} and Eq.~\eqref{eq_stabs_connected_4} do exist. The cases considered in this and the following section are lengthy as two special neighbors have to be chosen to make all other stabilizers commute with the measured parities via multiplication. This is necessary since the stabilizers in \cref{eq_stabs_connected_1,eq_stabs_connected_2} commute with both measured parities.

All stabilizers in Eq.~\eqref{eq_stabs_connected_3} as well as Eq.~\eqref{eq_stabs_connected_5} do not commute with the first measured parity $X_AZ_B$. We chose a special neighbor $a_*\in N(A)\setminus B\setminus N(B)$, and we multiply all the other stabilizers in \cref{eq_stabs_connected_3,eq_stabs_connected_5} with $S_{a_*}$. Furthermore, all stabilizers in \cref{eq_stabs_connected_4,eq_stabs_connected_5} do not commute with the second parity $Z_AX_B$ (also after multiplication with $S_{a_*}$). Therefore, we chose a special neighbor $b_* \in N(B)\setminus A\setminus N(A)$ and multiply $S_{b_*}$ on all the other stabilizers that anti-commute with $Z_AX_B$. Dropping the non-measurable stabilizers $S_{a_*}, S_{b_*}$ yields the stabilizers after the fusion:
\begin{align}
    &Z_{a_*}Z_{N(A)\setminus a_*}\\
    &Z_{b_*} Z_{N(B)\setminus b_*}\\
    &\forall a_i \in N(A)\setminus N(B)\setminus B\setminus a_*: X_{a_i}X_{a_*}Z_{N(a_i)\Delta N(a_*)}\\
    &\forall b_i \in N(B)\setminus N(A)\setminus A\setminus b_*: X_{b_i}X_{b_*}Z_{N(b_i)\Delta N(b_*)}\\
    &\forall c_i \in N(A)\cap N(B): X_{c_i}X_{a_*}X_{b_*}Z_{N(c_i)\Delta N(a_*)\Delta N(b_*)}\\
    &\forall d_i \notin N(B)\cup N(A): X_{d_i}Z_{N(d_i)}
\end{align}
To obtain a graph state, we apply two single-qubit Hadamard gates $H_{a_*}$, $H_{b_*}$. Using the definitions $K^*_{a_i}:=\left(N(a_i)\Delta N(a_*)\right)\setminus\{a_i,a_*,b_*\}$, $L^*_{b_i}:=\left(N(b_i)\Delta N(b_*)\right)\setminus\{b_i,a_*,b_*\}$, $M^*_{c_i}:=\left(N(c_i)\Delta N(a_*)\Delta N(b_*)\right)\setminus\{c_i,a_*,b_*\}$ yields the following stabilizers:
\begin{align}
    &X_{a_*}Z_{N(A)\setminus a_*}\label{eq:XZZX_multipl_1}\\
    &X_{b_*} Z_{N(B)\setminus b_*}\label{eq:XZZX_multipl_2}\\
    &\forall a_i \in N(A)\setminus N(B)\setminus B\setminus a_*:\notag\\
    &\hspace{0.5cm}\textbf{if: } a_i\notin N(a_*):\notag\\
    &\hspace{1.0cm}\textbf{if: } b_*\notin N(a_i)\Delta N(a_*): X_{a_i}Z_{a_*}Z_{K^*_{a_i}}\\
    &\hspace{1.0cm}\textbf{else: } X_{a_i}Z_{a_*}X_{b_*}Z_{K^*_{a_i}}\\
    &\hspace{0.5cm}\textbf{else: }\notag\\
    &\hspace{1.0cm}\textbf{if: } b_*\notin N(a_i)\Delta N(a_*): Y_{a_i}Y_{a_*}Z_{K^*_{a_i}}\\
    &\hspace{1.0cm}\textbf{else: } Y_{a_i}Y_{a_*}X_{b_*}Z_{K^*_{a_i}}\label{eq:XZZX_example_multip_1}\\
    &\forall b_i \in N(B)\setminus N(A)\setminus A\setminus b_*:\notag\\
    &\hspace{0.5cm}\textbf{if: } b_i\notin N(b_*):\notag\\
    &\hspace{1.0cm}\textbf{if: } a_*\notin N(b_i)\Delta N(b_*): X_{b_i}Z_{b_*}Z_{L^*_{b_i}}\\
    &\hspace{1.0cm}\textbf{else: } X_{b_i}X_{a_*}Z_{b_*}Z_{L^*_{b_i}}\\
    &\hspace{0.5cm}\textbf{else: }\notag\\
    &\hspace{1.0cm}\textbf{if: } a_*\notin N(b_i)\Delta N(b_*): Y_{b_i}Y_{b_*}Z_{L^*_{b_i}}\\
    &\hspace{1.0cm}\textbf{else: } Y_{b_i}X_{a_*}Y_{b_*}Z_{L^*_{b_i}}\\
    &\forall c_i \in N(A)\cap N(B):\notag\\
    &\hspace{0.5cm}\textbf{if: }\left(c_i\notin N(a_*)\land c_i\notin N(b_*)\land b_*\notin N(a_*)\right)\lor\left(c_i\in N(a_*)\land c_i\in N(b_*)\land b_*\in N(a_*)\right):\notag\\
    &\hspace{1.0cm} X_{c_i}Z_{a_*}Z_{b_*}Z_{M^*_{c_i}}\\
    &\hspace{0.5cm}\textbf{elif: }\left(c_i\in N(a_*)\land c_i\notin N(b_*)\land b_*\notin N(a_*)\right)\lor\left(c_i\notin N(a_*)\land c_i\in N(b_*)\land b_*\in N(a_*)\right):\notag\\
    &\hspace{1.0cm} Y_{c_i}Y_{a_*}Z_{b_*}Z_{M^*_{c_i}}\\
    &\hspace{0.5cm}\textbf{elif: }\left(c_i\notin N(a_*)\land c_i\in N(b_*)\land b_*\notin N(a_*)\right)\lor\left(c_i\in N(a_*)\land c_i\notin N(b_*)\land b_*\in N(a_*)\right):\notag\\
    &\hspace{1.0cm} Y_{c_i}Z_{a_*}Y_{b_*}Z_{M^*_{c_i}}\\
    &\hspace{0.5cm}\textbf{elif} \left(c_i\notin N(a_*)\land c_i\notin N(b_*)\land b_*\in N(a_*)\right)\lor\left(c_i\in N(a_*)\land c_i\in N(b_*)\land b_*\notin N(a_*)\right): \notag\\
    &\hspace{1.0cm} X_{c_i}Y_{a_*}Y_{b_*}Z_{M^*_{c_i}}\\
    &\forall d_i \notin N(B)\cup N(A):\notag\\
    &\hspace{0.5cm}\textbf{if } d_i\notin N(a_*) \land d_i\notin N(b_*): X_{d_i}Z_{N(d_i)}\\
    &\hspace{0.5cm}\textbf{elif } d_i\in N(a_*) \land d_i\notin N(b_*): X_{d_i}Z_{N(d_i)\setminus a_*}X_{a_*}\\
    &\hspace{0.5cm}\textbf{elif } d_i\notin N(a_*) \land d_i\in N(b_*): X_{d_i}Z_{N(d_i)\setminus b_*}X_{b_*}\\
    &\hspace{0.5cm}\textbf{elif } d_i\in N(a_*) \land d_i\in N(b_*): X_{d_i}Z_{N(d_i)\setminus\{a_*,b_*\}}X_{a_*}X_{b_*}
\end{align}
By suitable multiplication with the stabilizers in Eq.~\eqref{eq:XZZX_multipl_1} and/or Eq.~\eqref{eq:XZZX_multipl_2}, these equations are transformed into the standard representation of the stabilizer generators of a graph state. For instance: as the stabilizers in Eq.~\eqref{eq:XZZX_example_multip_1} have support on both $a_*, b_*$ with Pauli matrices that are not Pauli-$Z$, they are multiplied by both Eq.~\eqref{eq:XZZX_multipl_1} and Eq.~\eqref{eq:XZZX_multipl_2}. The stabilizers of the obtained graph state are:
\begin{align}
    &X_{a_*}Z_{N(A)\setminus a_*}\\
    &X_{b_*} Z_{N(B)\setminus b_*}\\
    &\forall a_i \in N(A)\setminus N(B)\setminus B\setminus a_*:\notag\\
    &\hspace{0.5cm}\textbf{if: } a_i\notin N(a_*):\notag\\
    &\hspace{1.0cm}\textbf{if: } b_*\notin N(a_i)\Delta N(a_*): X_{a_i}Z_{a_*}Z_{K^*_{a_i}}\\
    &\hspace{1.0cm}\textbf{else: } X_{a_i}Z_{a_*}Z_{(N(B)\setminus b_*)\Delta K^*_{a_i}}\\
    &\hspace{0.5cm}\textbf{else: }\notag\\
    &\hspace{1.0cm}\textbf{if: } b_*\notin N(a_i)\Delta N(a_*): X_{a_i}Z_{a_*}Z_{(N(A)\setminus\{a_*,a_i\})\Delta K^*_{a_i}}\\
    &\hspace{1.0cm}\textbf{else: } X_{a_i}Z_{a_*}Z_{(N(A)\setminus\{a_*,a_i\})\Delta (N(B)\setminus b_*)\Delta K^*_{a_i}}\label{eq_xzzx_example_1}\\
    &\forall b_i \in N(B)\setminus N(A)\setminus A\setminus b_*:\notag\\
    &\hspace{0.5cm}\textbf{if: } b_i\notin N(b_*):\notag\\
    &\hspace{1.0cm}\textbf{if: } a_*\notin N(b_i)\Delta N(b_*): X_{b_i}Z_{b_*}Z_{L^*_{b_i}}\\
    &\hspace{1.0cm}\textbf{else: } X_{b_i}Z_{b_*}Z_{(N(A)\setminus a_*)\Delta L^*_{b_i}}\\
    &\hspace{0.5cm}\textbf{else: }\notag\\
    &\hspace{1.0cm}\textbf{if: } a_*\notin N(b_i)\Delta N(b_*): X_{b_i}Z_{b_*}Z_{(N(B)\setminus\{b_*,b_i\})\Delta L^*_{b_i}}\\
    &\hspace{1.0cm}\textbf{else: } X_{b_i}Z_{b_*}Z_{(N(A)\setminus a_*)\Delta (N(B)\setminus\{b_*,b_i\})\Delta L^*_{b_i}}\\
    &\forall c_i \in N(A)\cap N(B):\notag\\
    &\hspace{0.5cm}\textbf{if: }\left(c_i\notin N(a_*)\land c_i\notin N(b_*)\land b_*\notin N(a_*)\right)\lor\left(c_i\in N(a_*)\land c_i\in N(b_*)\land b_*\in N(a_*)\right):\notag\\
    &\hspace{1.0cm} X_{c_i}Z_{a_*}Z_{b_*}Z_{M^*_{c_i}}\\
    &\hspace{0.5cm}\textbf{elif: }\left(c_i\in N(a_*)\land c_i\notin N(b_*)\land b_*\notin N(a_*)\right)\lor\left(c_i\notin N(a_*)\land c_i\in N(b_*)\land b_*\in N(a_*)\right):\notag\\
    &\hspace{1.0cm} X_{c_i}Z_{a_*}Z_{b_*}Z_{(N(A)\setminus\{a_*,c_i\})\Delta M^*_{c_i}}\label{eq_xzzx_example_2}\\
    &\hspace{0.5cm}\textbf{elif: }\left(c_i\notin N(a_*)\land c_i\in N(b_*)\land b_*\notin N(a_*)\right)\lor\left(c_i\in N(a_*)\land c_i\notin N(b_*)\land b_*\in N(a_*)\right):\notag\\
    &\hspace{1.0cm} X_{c_i}Z_{a_*}Z_{b_*}Z_{(N(B)\setminus\{b_*,c_i\})\Delta M^*_{c_i}}\\
    &\hspace{0.5cm}\textbf{elif} \left(c_i\notin N(a_*)\land c_i\notin N(b_*)\land b_*\in N(a_*)\right)\lor\left(c_i\in N(a_*)\land c_i\in N(b_*)\land b_*\notin N(a_*)\right): \notag\\
    &\hspace{1.0cm} X_{c_i}Z_{a_*}Z_{b_*}Z_{(N(A)\setminus a_*)\Delta (N(B)\setminus b_*)\Delta M^*_{c_i}}\label{eq_xzzx_example_3}\\
    &\forall d_i \notin N(B)\cup N(A):\notag\\
    &\hspace{0.5cm}\textbf{if } d_i\notin N(a_*) \land d_i\notin N(b_*): X_{d_i}Z_{N(d_i)}\\
    &\hspace{0.5cm}\textbf{elif } d_i\in N(a_*) \land d_i\notin N(b_*): X_{d_i}Z_{N(d_i)\Delta N(A)}\\
    &\hspace{0.5cm}\textbf{elif } d_i\notin N(a_*) \land d_i\in N(b_*): X_{d_i}Z_{N(d_i)\Delta N(B)}\\
    &\hspace{0.5cm}\textbf{else}: X_{d_i}Z_{N(d_i)\Delta N(A)\Delta N(B)}
\end{align}
As before, all these equations correspond to symmetric differences between neighborhoods from the original graph state. In the case of Eq.~\eqref{eq_xzzx_example_3}, five different graph neighborhoods are involved, which is the maximum number for all the considered graph transformation rules.

\subsubsection{$N(A)\setminus B\setminus N(B)\neq\emptyset\land N(B)\setminus A\setminus N(A)=\emptyset$}
In this case, the neighborhood $N(B)\setminus A$ is fully contained in $N(A)$ with a finite number of neighbors in $N(B)\setminus A$. There are stabilizers of the type $S_{a_i}, S_{c_i}$ in \cref{eq_stabs_connected_3,eq_stabs_connected_5} but no stabilizers of the type $S_{b_i}$ in Eq.~\eqref{eq_stabs_connected_4}. Both $S_{a_i}$ and $S_{c_i}$ anticommute with the measured parity $X_AZ_B$ and a stabilizer corresponding to a special neighbor $S_{a_*}$ is multiplied on all these equations. Eq.~\eqref{eq_stabs_connected_5} changes to $S_{c_i}S_{a_*}$ which anticommutes with the second measured parity ($Z_AX_B$). Therefore, we chose a second special neighbor $c_*$ and multiply $S_{c_*}$ on all other stabilizers of type $S_{c_i}S_{a_*}$. After applying $H_{a_*}, H_{c_*}$, a derivation similar to the previous section yields:
\begin{align}
    &X_{a_*}Z_{N(A)\setminus N(B) \setminus a_*}\\
    &X_{c_*} Z_{N(B)\setminus c_*} \label{eq:XZZX_connected_last_case_c_special}\\
    &\forall a_i \in N(A)\setminus N(B)\setminus B\setminus a_*:\notag\\
    &\hspace{0.5cm}\textbf{if: } a_i\notin N(a_*):\notag\\
    &\hspace{1.0cm}\textbf{if: } c_*\notin N(a_i)\Delta N(a_*): X_{a_i}Z_{a_*}Z_{K^*_{a_i}}\\
    &\hspace{1.0cm}\textbf{else: } X_{a_i}Z_{a_*}Z_{(N(B)\setminus c_*)\Delta K^*_{a_i}}\\
    &\hspace{0.5cm}\textbf{else: }\notag\\
    &\hspace{1.0cm}\textbf{if: } c_*\notin N(a_i)\Delta N(a_*): X_{a_i}Z_{a_*}Z_{(N(A)\setminus\{N(B),a_*,a_i\})\Delta K^*_{a_i}}\\
    &\hspace{1.0cm}\textbf{else: } X_{a_i}Z_{a_*}Z_{(N(A)\setminus\{N(B),a_*,a_i\})\Delta (N(B)\setminus c_*)\Delta K^*_{a_i}}\\
    &\forall c_i \in N(A)\cap N(B)\setminus c_*:\notag\\
    &\hspace{0.5cm}\textbf{if: } c_i\notin N(c_*):\notag\\
    &\hspace{1.0cm}\textbf{if: } a_*\notin N(c_i)\Delta N(c_*): X_{c_i}Z_{c_*}Z_{L^*_{c_i}}\\
    &\hspace{1.0cm}\textbf{else: } X_{c_i}Z_{c_*}Z_{(N(A)\setminus N(B)\setminus a_*)\Delta L^*_{c_i}}\\
    &\hspace{0.5cm}\textbf{else: }\notag\\
    &\hspace{1.0cm}\textbf{if: } a_*\notin N(c_i)\Delta N(c_*): X_{c_i}Z_{c_*}Z_{(N(B)\setminus\{c_*,c_i\})\Delta L^*_{c_i}}\\
    &\hspace{1.0cm}\textbf{else: } X_{c_i}Z_{c_*}Z_{(N(A)\setminus N(B)\setminus a_*)\Delta (N(B)\setminus\{c_*,c_i\})\Delta L^*_{c_i}}\\
    &\forall d_i \notin N(A)\cup N(B):\notag\\
    &\hspace{0.5cm}\textbf{if } d_i\notin N(a_*) \land d_i\notin N(c_*): X_{d_i}Z_{N(d_i)}\\
    &\hspace{0.5cm}\textbf{elif } d_i\in N(a_*) \land d_i\notin N(c_*): X_{d_i}Z_{N(d_i)\Delta (N(A)\setminus N(B))}\\
    &\hspace{0.5cm}\textbf{elif } d_i\notin N(a_*) \land d_i\in N(c_*): X_{d_i}Z_{N(d_i)\Delta N(B)}\\
    &\hspace{0.5cm}\textbf{else}: X_{d_i}Z_{N(d_i)\Delta (N(A)\setminus N(B))\Delta N(B)}
\end{align}
With $K^*_{a_i}:=\left(N(a_i)\Delta N(a_*)\right)\setminus\{a_i,a_*,c_*\}$ and $L^*_{c_i}:=\left(N(c_i)\Delta N(c_*)\right)\setminus\{c_i,a_*,c_*\}$.

\subsubsection{$N(B)\setminus A\setminus N(A)\neq\emptyset\land N(A)\setminus B\setminus N(B)=\emptyset$}
This case is analogous to the previous section by interchanging the labels $A, B$.

\subsection{$X_AY_B\land Y_AZ_B$ ($A$, $B$ connected)}
In Section~\ref{sec:xyyz_connect}, we have considered the parity measurement $X_AY_B$ $\land$ $Y_AZ_B$ when the two fusion qubits $A, B$ are not connected. Here we derive the corresponding rules when $A, B$ are connected starting from \cref{eq_stabs_connected_1,eq_stabs_connected_2,eq_stabs_connected_3,eq_stabs_connected_4,eq_stabs_connected_5,eq_stabs_connected_6}. The stabilizer $S_A$ in Eq~\eqref{eq_stabs_connected_1} anticommutes with $X_AY_B$. The same applies to $S_{a_i}$ (Eq.~\eqref{eq_stabs_connected_3}) and $S_{b_i}$ (Eq.~\eqref{eq_stabs_connected_4}) and we replace $S_{a_i}$ by $S_{a_i}S_A$, $S_{b_i}$ by $S_{b_i}S_A$, and drop $S_A$. The remaining stabilizers, which all commute with $X_AY_B$, are:
\begin{align}
    &S_B = Z_{N(B)\setminus A}X_B Z_A\label{eq:xyyz_conn_1}\\
    &\forall a_i \in N(A)\setminus B\setminus N(B): S_{a_i}=Y_{a_i}Z_{N(a_i)\setminus A}Z_{N(A)\setminus a_i\setminus B}Y_AZ_B\label{eq:xyyz_conn_2}\\
    &\forall b_i \in N(B)\setminus A\setminus N(A): S_{b_i}=X_{b_i}Z_{N(b_i)\Delta N(A)}X_A\label{eq_xyyz_conn_3}\\
    &\forall c_i \in N(A)\cap N(B): S_{c_i}=X_{c_i}Z_{N(c_i)\setminus A \setminus B}Z_AZ_B\label{eq_xyyz_conn_4}\\
    &\forall d_i \notin N(B)\cup N(A): X_{d_i}Z_{N(d_i)}\label{eq:xyyz_conn_5}
\end{align}
where the stabilizers in \cref{eq_xyyz_conn_3,eq_xyyz_conn_4} anticommute with the second parity $Y_AZ_B$. As before, different cases are considered since some types of anticommuting stabilizers may not exist.

\subsubsection{$N(B)\setminus A\setminus N(A)=\emptyset\land N(B)\cap N(A)=\emptyset\land N(A)\setminus B\setminus N(B)\neq\emptyset$}
In this case, no stabilizer from \cref{eq_xyyz_conn_3,eq_xyyz_conn_4} exists and the stabilizer in Eq.~\eqref{eq:xyyz_conn_1} is equal to $\mathds{1}$ after the fusion. $S_{a_i}$ in Eq.~\eqref{eq:xyyz_conn_2} and $S_{d_i}$ in Eq.~\eqref{eq:xyyz_conn_5} are the only remaining stabilizers, and these commute with both measured parities. Applying the gate $R$ on the qubits of type $S_{a_i}$ and dropping the measured qubits $A, B$ yields the stabilizers of a graph state:
\begin{align}
    &\forall a_i \in N(A)\setminus B\setminus N(B): S_{a_i}=X_{a_i}Z_{N(a_i)\Delta (N(A)\setminus a_i)}\\
    &\forall d_i \notin N(B)\cup N(A): X_{d_i}Z_{N(d_i)}
\end{align}

\subsubsection{$N(B)\setminus A\setminus N(A)\neq\emptyset$}
In this case, there is at least one stabilizer of type $S_{b_i}$ in Eq.~\eqref{eq_xyyz_conn_3} that anticommutes with the second parity $Y_AZ_B$. We choose a special neighbor $b_*\in N(B)\setminus A\setminus N(A)$ and multiply $S_{b_*}$ on all other stabilizers that anticommute with $Y_AZ_B$. Dropping the measured qubits $A, B$ and writing the support on $b_*$ explicitly yields:
\begin{align}
    &Z_{b_*}Z_{N(B)\setminus b_*}\\
    &\forall a_i \in N(A)\setminus B\setminus N(B):\notag\\
    &\hspace{0.5cm}\textbf{if } a_i\notin N(b_*): Y_{a_i}Z_{N(a_i)\Delta (N(A)\setminus a_i)}\\
    &\hspace{0.5cm}\textbf{else }: Y_{a_i}Z_{b_*}Z_{(N(a_i)\setminus b_*)\Delta (N(A)\setminus a_i)}\\
    &\forall b_i \in N(B)\setminus A\setminus b_*\setminus N(A):\notag\\
    &\hspace{0.5cm}\textbf{if } b_i\notin N(b_*):X_{b_i}X_{b_*}Z_{N(b_i)\Delta N(b_*)}\\
    &\hspace{0.5cm}\textbf{else }: Y_{b_i}Y_{b_*}Z_{(N(b_i)\setminus b_*)\Delta (N(b_*)\setminus b_i)}\\
    &\forall c_i \in N(A)\cap N(B):\notag\\
    &\hspace{0.5cm}\textbf{if } c_i\notin N(b_*): Y_{c_i}X_{b_*}Z_{N(c_i)}Z_{N(b_*)\Delta (N(A)\setminus c_i)}\\
    &\hspace{0.5cm}\textbf{else }: X_{c_i}Y_{b_*}Z_{N(c_i)\setminus b_*}Z_{N(b_*)\Delta N(A)}\\
    &\forall d_i \notin N(B)\cup N(A):\notag\\
    &\hspace{0.5cm}\textbf{if } d_i\notin N(b_*): X_{d_i}Z_{N(d_i)}\\
    &\hspace{0.5cm}\textbf{else }: X_{d_i}Z_{N(d_i)\Delta N(B)}
\end{align}
We multiply stabilizers containing $Z_{b_*}$ or $Y_{b_*}$ with $S_B$ in Eq.~\eqref{eq:xyyz_conn_1}. Then we apply the gate $H_{b_*}$. Furthermore, we apply the gate $R$ on all qubits $a_i \in N(A)\setminus B\setminus N(B)$ and on all qubits of type $c_i \in N(A)\cap N(B)$. This yields the following graph state stabilizers:
\begin{align}
    &X_{b_*}Z_{N(B)\setminus b_*}\\
    &\forall a_i \in N(A)\setminus B\setminus N(B):\notag\\
    &\hspace{0.5cm}\textbf{if } a_i\notin N(b_*): X_{a_i}Z_{N(a_i)\Delta (N(A)\setminus a_i)}\\
    &\hspace{0.5cm}\textbf{else: }X_{a_i}Z_{N(a_i)\Delta N(B)\Delta (N(A)\setminus a_i)}\\
    &\forall b_i \in N(B)\setminus A\setminus b_*\setminus N(A):\notag\\
    &\hspace{0.5cm}\textbf{if } b_i\notin N(b_*):X_{b_i}Z_{b_*}Z_{N(b_i)\Delta N(b_*)}\\
    &\hspace{0.5cm}\textbf{else }: X_{b_i}Z_{b_*}Z_{N(b_i)\Delta N(b_*)\Delta N(B)}\\
    &\forall c_i \in N(A)\cap N(B):\notag\\
    &\hspace{0.5cm}\textbf{if } c_i\notin N(b_*): X_{c_i}Z_{b_*}Z_{N(c_i)\Delta N(b_*)\Delta (N(A)\setminus c_i)} \label{eq:XYYZ_connected_first_case_ci_not_in}\\
    &\hspace{0.5cm}\textbf{else: } X_{c_i}Z_{b_*}Z_{N(c_i)\Delta N(A)\Delta N(B)\Delta (N(b_*)\setminus c_i)}\label{eq:XYYZ_connected_first_case_ci_in}\\
    &\forall d_i \notin N(B)\cup N(A):\notag\\
    &\hspace{0.5cm}\textbf{if } d_i\notin N(b_*): X_{d_i}Z_{N(d_i)}\\
    &\hspace{0.5cm}\textbf{else: } X_{d_i}Z_{N(d_i)\Delta N(B)}
\end{align}
\subsubsection{$N(B)\setminus A\setminus N(A)=\emptyset\land N(B)\cap N(A)\neq\emptyset$}
Here, no neighbor $b_*\in N(B)\setminus A\setminus N(A)$ exists, and therefore we choose a special neighbor $c_*\in N(B)\cap N(A)$ and multiply $S_{c_*}$ on all the other stabilizers that anticommute with $Y_AZ_B$. A procedure that is very similar to the previous section yields the following stabilizers after applying the Hadamard gate $H_{c_*}$ and the gates $R_{a_i}$ on all qubits $a_i \in N(A)\setminus B\setminus N(B)$:
\begin{align}
    &X_{c_*}Z_{N(B)\setminus c_*}\\
    &\forall a_i \in N(A)\setminus B\setminus N(B):\notag\\
    &\hspace{0.5cm}\textbf{if } a_i\notin N(c_*): X_{a_i}Z_{N(a_i)\Delta N(B)\Delta (N(A)\setminus a_i)}\\
    &\hspace{0.5cm}\textbf{else: }X_{a_i}Z_{N(a_i)\Delta (N(A)\setminus a_i)}\\
    &\forall c_i \in (N(A)\cap N(B))\setminus c_*:\notag\\
    &\hspace{0.5cm}\textbf{if } c_i\notin N(c_*): X_{c_i}Z_{c_*}Z_{N(c_i)\Delta N(c_*)}\\
    &\hspace{0.5cm}\textbf{else: }X_{c_i}Z_{c_*}Z_{N(c_i)\Delta N(c_*)\Delta N(B)}\\
    &\forall d_i \notin N(B)\cup N(A):\notag\\
    &\hspace{0.5cm}\textbf{if } d_i\notin N(c_*): X_{d_i}Z_{N(d_i)}\\
    &\hspace{0.5cm}\textbf{else: } X_{d_i}Z_{N(d_i)\Delta N(B)} \label{eq:xzzx_C_last_case_di_not_in}
\end{align}

\subsection{$X_AX_B\land Z_AZ_B$ ($A$, $B$ connected)}
The graph transformations are identical to the case in which the fusion qubits $A, B$ are not connected. The reason is the following: first, \cref{eq_stabs_all_1,eq_stabs_all_2,eq_stabs_all_3,eq_stabs_all_4,eq_stabs_all_5,eq_stabs_all_6} and \cref{eq_stabs_connected_1,eq_stabs_connected_2,eq_stabs_connected_3,eq_stabs_connected_4,eq_stabs_connected_5,eq_stabs_connected_6} are identical when disregarding the measured qubits $A, B$. Second, the procedure to make stabilizers (anticommuting with the measured parities) commute can be chosen identically: multiply $S_A$ on $S_B$ and if $(N(A)\setminus B)\Delta(N(B)\setminus A)\neq\emptyset$, choose a special neighbor in $(N(A)\setminus B)\Delta(N(B)\setminus A)$ and multiply the corresponding stabilizer with all other ones that anticommute with the second measured parity. If $(N(A)\setminus B)\Delta(N(B)\setminus A)=\emptyset$, $S_AS_B=\mathds{1}$ can be removed and the fusion is just removing both qubits from the graph. Thus, the graph transformation is the same as in Section~\ref{sec:XXZZ_detached}.

\subsection{$Y_AZ_B\land Z_AY_B$ ($A$, $B$ connected)}
\label{subsec:YZ_ZY-connected}
In this section, we derive the update rules for the parity measurement $Y_AZ_B \land Z_AY_B$ when the fusion qubits $A$ and $B$ are connected. The update rules are realized by first multiplying Eq.~\eqref{eq_stabs_connected_1} on \cref{eq_stabs_connected_3,eq_stabs_connected_5} followed by multiplying Eq.~\eqref{eq_stabs_connected_2} on \cref{eq_stabs_connected_4,eq_stabs_connected_5}, and finally applying $R$ on the qubits $a_i\in N(A)\setminus N(B)\setminus B$ and $b_i\in N(B)\setminus N(A)\setminus A$ (\cref{eq_stabs_connected_3,eq_stabs_connected_4}). Omitting the fusion qubits yields the stabilizers of the transformed graph state:
\begin{align}
    &\forall a_i \in N(A)\setminus N(B)\setminus B: X_{a_i}Z_{N(a_i)\Delta N(A) \setminus a_i} \label{eq:YZZY_C_ai} \\
    &\forall b_i \in N(B)\setminus N(A)\setminus A: X_{b_i}Z_{N(b_i)\Delta N(B) \setminus b_i} \label{eq:YZZY_C_bi} \\
    &\forall c_i \in N(A)\cap N(B):  
    X_{c_i}Z_{N(c_i)\Delta N(B)\Delta N(A)} \label{eq:YZZY_C_ci}\\
    &\forall d_i \notin N(B)\cup N(A): 
    X_{d_i}Z_{N(d_i)} \label{eq:YZZY_C_di}
\end{align}

\subsection{$Y_AX_B\land X_AY_B$ ($A$, $B$ connected)}
The graph transformations are identical to the case in which the fusion qubits $A, B$ are not connected (see Appendix~\ref{sec:xyyx_detached}). The reason is that \cref{eq_stabs_all_1,eq_stabs_all_2,eq_stabs_all_3,eq_stabs_all_4,eq_stabs_all_5,eq_stabs_all_6} and \cref{eq_stabs_connected_1,eq_stabs_connected_2,eq_stabs_connected_3,eq_stabs_connected_4,eq_stabs_connected_5,eq_stabs_connected_6} are identical up to the measured qubits $A, B$, and in both cases, the stabilizers $S_A, S_B$ can be multiplied on all stabilizers in $N(A)\Delta N(B)$ to make all other stabilizers commute with the measured parties. Therefore, \cref{eq_YXXY_D_in_symmetric_difference,eq_YXXY_D_not_in_symmetric_difference} also represent the graph transformation for connected fusion qubits.

\section{Graph transformation by single-qubit measurements}
\label{sec:single_measure}
For completeness, we also show here the graph transformations corresponding to single-qubit measurements. These transformations can also be found in Refs.~\cite{Hein2004,Hein2006,Anders2006}. Before the measurement of qubit $A$, the stabilizers are:
\begin{align}
    &S_A=X_AZ_{N(A)}\label{eq_1measure_1}\\
    &\forall a_i \in N(A): S_{a_i}=X_{a_i}Z_{N(a_i)\setminus A}Z_A\label{eq_1measure_2}\\
    &\forall d_i \notin \left(N(A)\cup A\right): S_{d_i}=X_{d_i}Z_{N(d_i)}
\end{align}
When measuring qubit $A$ in the $Z$-basis, the only stabilizer that anticommutes with the measurement is the one in Eq.~\eqref{eq_1measure_1}. Up to local $Z$-gates applied to $N(i)$ upon the measurement outcome $1$, the $Z$-basis measurement therefore corresponds to removing the measured qubit from the graph. Measurements in the $X$- and $Y$-basis can be represented by local graph complementations~\cite{Hein2006} or symmetric differences of local neighborhoods in the graph~\cite{Anders2006} in combination with removing the measured qubits.

\subsection{single-qubit $Y$-basis measurement}
\begin{figure*}
\includegraphics[width=1.0\columnwidth]{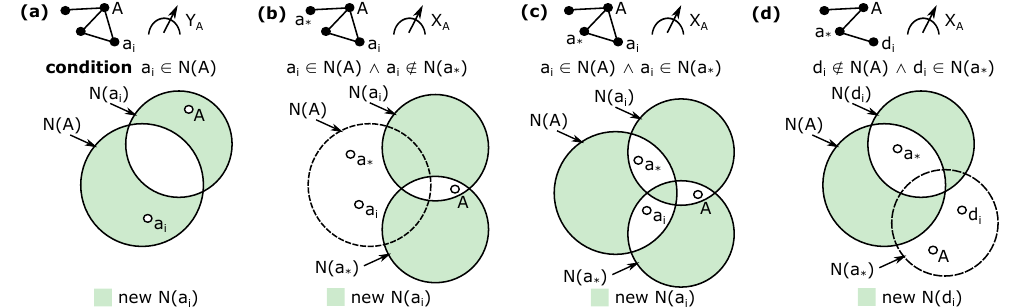}
\caption{\label{fig:appendix_single}Venn diagrams representing the neighborhood transformations for single-qubit measurements. \textbf{(a)} Effect of measuring $Y_A$ (see Eq.~\eqref{eq:single_0}). \textbf{(b, c, d)} Effect of measuring $X_A$ for the cases given by Eq.~\eqref{eq:single_1}, Eq.~\eqref{eq:single_2}, and Eq.~\eqref{eq:single_3}, respectively.}
\end{figure*}
The \cref{eq_1measure_1,eq_1measure_2} do both anticommute with $Y_A$. When measuring $Y_A$, the remaining stabilizers can be obtained by multiplying Eq.~\eqref{eq_1measure_1} on all stabilizers in Eq.~\eqref{eq_1measure_2} and dropping Eq.~\eqref{eq_1measure_1} afterward yields the following:
\begin{align}
    &\forall a_i \in N(A): Y_{a_i}Z_{N(a_i) \Delta\left(N(A)\setminus a_i\right)}\\
    &\forall d_i \notin N(A): X_{d_i}Z_{N(d_i)}
\end{align}
where, as before, we have not explicitly subtracted the qubit $A$ from $N(a_i)$ as it has been measured and thus is not part of the graph anymore. After applying the gate $R$ on all qubits $a_i$:
\begin{align}
    &\forall a_i \in N(A): X_{a_i}Z_{N(a_i)\Delta\left(N(A)\setminus a_i\right)}\label{eq:single_0}\\
    &\forall d_i \notin N(A): X_{d_i}Z_{N(d_i)}
\end{align}
This corresponds to adding (modulo-two) $N(A)$ on the neighborhoods of all qubits within $N(A)$, respectively, a local complementation of the neighborhood $N(A)$~\cite{Hein2004}.

\subsection{single-qubit $X$-basis measurement}
The stabilizer in Eq.~\eqref{eq_1measure_1} commutes with $X_A$ but the stabilizers in Eq.~\eqref{eq_1measure_2} anticommute with it. We chose a special neighbor $a_*$ and multiply $S_{a_*}$ on the other stabilizers in Eq.~\eqref{eq_1measure_2}. Applying $H_{a_*}$ and removing the measured qubit $A$ from Eq.~\eqref{eq_1measure_1} results in:
\begin{align}
    &X_{a_*}Z_{N(A)\setminus a_*}\\
    &\forall a_i \in N(A)\setminus a_*:\notag\\
    &\hspace{0.5cm}\textbf{if } a_i\notin N(a_*): X_{a_i}Z_{a_*}Z_{N(a_i)\Delta N(a_*)}\label{eq:single_1}\\
    &\hspace{0.5cm}\textbf{else }: X_{a_i}Z_{a_*}Z_{N(a_i)\Delta N(a_*)\Delta N(A)}\label{eq:single_2}\\
    &\forall d_i \notin N(A):\notag\\
    &\hspace{0.5cm}\textbf{if } d_i\notin N(a_*): X_{d_i}Z_{N(d_i)}\\
    &\hspace{0.5cm}\textbf{else }: X_{d_i}Z_{N(d_i)\Delta N(A)}\label{eq:single_3}
\end{align}
The Venn diagrams corresponding to the graph transformations upon single-qubit $Y$- and $X$-basis measurements are shown in Fig.~\ref{fig:appendix_single}.

\section{Resource state generation}
\label{sec:res_generate}
\begin{figure*}
\includegraphics[width=1.0\textwidth]{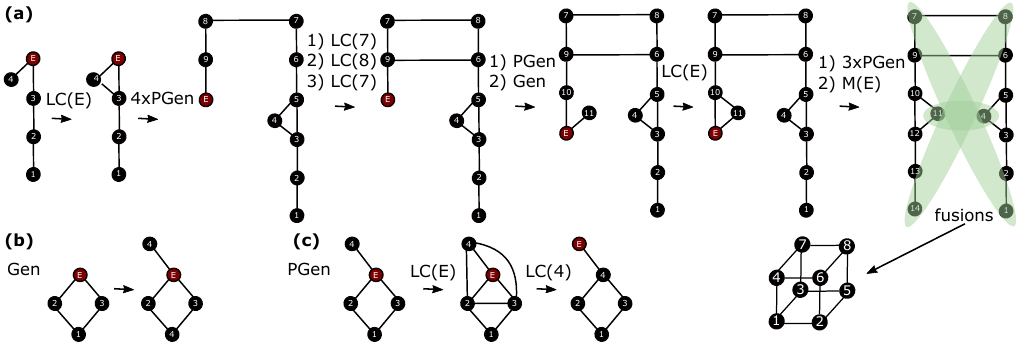}
\caption{\label{fig:Cube_generation}\textbf{(a)} Generation of the cubic graph state~\cite{Bell2022} using a single quantum emitter and the three fusion operations from Fig.~\ref{fig:fNet}(c) of the main text. \textbf{(b)} Explanation of the operation \textit{Gen}. A new photon attached to the quantum emitter, $E$, by a controlled-$Z$ gate is generated by optical excitation of the emitter followed by photon emission and a Hadamard gate on the emitted photon (see Refs.~\cite{Lindner2009, Tiurev2021, Paesani2022} for more details on the scheme). \textbf{(c)} Explanation of the operation \textit{P.Gen}. The operation is composed of generating a photonic qubit as in (b) followed by local complementation (LC) on the emitter and then LC on the newly generated photonic qubit.}
\end{figure*}

In the main text, we have shown a construction of the cube graph from the resource state in Fig.~\ref{fig:fNet} using three fusions. In Fig.~\ref{fig:Cube_generation}(a) we show how the used resource state can be deterministically generated by a single quantum emitter (see Refs.~\cite{Lindner2009,Tiurev2021,Li2022} for related schemes). The generation relies on three operations: 1) Photonic qubit generation (see operation \textit{Gen} in Fig.~\ref{fig:Cube_generation}(b)), 2) Local complementation (LC)~\cite{Hein2006}\footnote{Applying LC on a node $q$ of a graph $G$ complements the induced subgraph $G[N(q)]$ (the subgraph with vertices from the neighborhood $N(q)$, connected by edges from $G$). Local complementation on qubit $q$ can be implemented by the single-qubit Clifford gates $\exp\left(-i\frac{\pi}{4}X_q\right)\exp\left(i\frac{\pi}{4}Z\right)_{N(q)}$~\cite{Hein2004, Hein2006}}, and 3) single-qubit measurements in the $Z$-basis. In a graph picture, the generation of a photonic qubit corresponds to adding an edge between the emitter node and the generated photonic qubit node~\cite{Lindner2009,Paesani2022}, as shown in Fig.~\ref{fig:Cube_generation}(b). Furthermore, first generating a photonic qubit, then LC on the emitter followed by LC on the newly generated photonic qubit interchanges the emitter and the photonic qubit in the graph (see Fig.~\ref{fig:Cube_generation}(c)). This operation we call \textit{Push photonic qubit generation} (\textit{P.Gen}). With these operations, we generate the required resource state as shown in Fig.~\ref{fig:Cube_generation}(a), followed by generating the cube graph using the three fusions. Without any fusions, we find by computing the so-called height function~\cite{Li2022} for all orderings of the photonic qubit generation that at least three interacting quantum emitters would be required to generate the cube graph deterministically.

\end{appendices}
\end{document}